\documentclass[aps,twocolumn,prx]{revtex4-1}
\usepackage{graphicx,amsmath, amsfonts,bm, color}

\usepackage{xcolor,hyperref}
\hypersetup{
   colorlinks,
   linkcolor={blue!50!black},
   citecolor={blue!50!black},
   urlcolor={blue!80!black}
}

\usepackage{enumitem}

\renewcommand{\=}{\!=\!}
\newcommand{\1}{^{\mbox{\tiny (1)}}}

\DeclareMathOperator{\sgn}{sgn}

\begin{document}

\title{The emergence of crack-like behavior of frictional rupture:\\ Edge singularity and energy balance}
\author{Fabian Barras$^{1}$}
\author{Michael Aldam$^{2}$}
\author{Thibault Roch$^{1}$}
\author{Efim A.~Brener$^{3,4}$}
\author{Eran Bouchbinder$^{2}$}
\thanks{eran.bouchbinder@weizmann.ac.il}
\author{Jean-Fran\c{c}ois Molinari$^{1}$}
\thanks{jean-francois.molinari@epfl.ch}
\affiliation{$^{1}$Civil Engineering Institute, Materials Science and Engineering Institute, Ecole Polytechnique F\'ed\'erale de Lausanne, Station 18, CH-1015 Lausanne, Switzerland\\
$^{2}$Chemical and Biological Physics Department, Weizmann Institute of Science, Rehovot 7610001, Israel\\
$^{3}$Peter Gr\"unberg Institut, Forschungszentrum J\"ulich, D-52425 J\"ulich, Germany\\
$^{4}$Institute for Energy and Climate Research, Forschungszentrum J\"ulich, D-52425 J\"ulich, Germany}

\begin{abstract}
The failure of frictional interfaces --- the process of frictional rupture --- is widely assumed to feature crack-like properties, with far-reaching implications for various disciplines, ranging from engineering tribology to earthquake physics. An important condition for the emergence of a crack-like behavior is the existence of stress drops in frictional rupture, whose basic physical origin has been recently elucidated. Here we show that for generic and realistic frictional constitutive relations, and once the necessary conditions for the emergence of an effective crack-like behavior are met, frictional rupture dynamics are approximately described by a crack-like, fracture mechanics energy balance equation. This is achieved by independently calculating the intensity of the crack-like singularity along with its associated elastic energy flux into the rupture edge region, and the frictional dissipation in the edge region. We further show that while the fracture mechanics energy balance equation provides an approximate, yet quantitative, description of frictional rupture dynamics, interesting deviations from the ordinary crack-like framework --- associated with non-edge-localized dissipation --- exist. Together with the recent results about the emergence of stress drops in frictional rupture, this work offers a comprehensive and basic understanding of why, how and to what extent frictional rupture might be viewed as an ordinary fracture process. Various implications are discussed.
\end{abstract}

\maketitle

\section{Background and motivation}
\label{sec:intro}

Rapid slip along interfaces separating bodies in frictional contact is mediated by the spatiotemporal dynamics of frictional rupture~\citep{Svetlizky2019,Scholz2002}, which is a fundamental process of prime importance for a broad range of physical systems. For example, it is responsible for squealing in car brake pads~\citep{Rhee1991}, for bowing on a violin string~\citep{Casado2017}, and for earthquakes along geological faults~\citep{Marone1998a,Ben-Zion2008,Ohnaka2013}, to name just a few well-known examples. A very powerful conceptual and quantitative framework to understand frictional dynamics in a wide variety of physical contexts is the analogy between frictional rupture and ordinary fracture/cracks.

This framework is extensively used to interpret and quantify geophysical observations~\citep{Abercrombie2005,Bizzarri2016}, as well as a broad spectrum of laboratory phenomena~\citep{Lu2010,Lu2010a,Noda2013a,Svetlizky2014,Bayart2015,Svetlizky2016,Rubino2017,Svetlizky2017a}. For example, a recent series of careful laboratory experiments~\citep{Svetlizky2014,Bayart2015,Svetlizky2016} demonstrated that when the analogy between frictional rupture and ordinary fracture holds, the dynamic propagation of laboratory earthquakes and their arrest can be quantitatively understood to an unprecedented degree~\citep{Kammer2015}. Yet, the fundamental physical origin and range of validity of the analogy between frictional rupture and ordinary fracture are not yet fully understood.

An important condition for the analogy to hold is the emergence of a finite and well-defined stress drop $\Delta\tau\=\tau_{\rm d}-\tau_{\rm res}$, the difference between the applied driving stress $\tau_{\rm d}$ and the residual stress $\tau_{\rm res}$, in frictional rupture. In a very recent paper~\citep{PartI} we showed that, contrary to widely adopted assumptions, the residual stress $\tau_{\rm res}$ is not a characteristic property of frictional interfaces. Rather, for rapid rupture $\tau_{\rm res}$ is shown to crucially depend on elastodynamic bulk effects --- in particular wave radiation from the frictional interface to the bulks surrounding it and long-range elastodynamic bulk interactions ---  and that the existence of a finite stress drop $\Delta\tau$, is a finite time effect, limited by the wave travel time in finite systems. Specifically, it has been shown that
\begin{equation}
\label{eq:SD}
\Delta\tau(\tau_{\rm d}) \simeq \frac{\mu}{2c_s}v^0_{\rm res}(\tau_{\rm d}) \ ,
\end{equation}
where $\mu$ is the shear modulus of the bulks surrounding the frictional interface, $c_s$ is the corresponding shear wave-speed and $v^0_{\rm res}$ is the theoretically predicted residual slip velocity behind the propagating rupture edge. $v^0_{\rm res}(\tau_{\rm d})$ is determined through the approximate equation $\tau_{\rm ss}(v^0_{\rm res})+\frac{\mu}{2c_s}v^0_{\rm res}\!\simeq\!\tau_{\rm d}$, once long-range elastodynamic contributions are omitted~\citep{PartI}, where $\tau_{\rm ss}(v)$ is the steady-state friction curve as a function of slip velocity $v$.

\begin{figure}[ht!]
  \centering
  \includegraphics[width=\columnwidth]{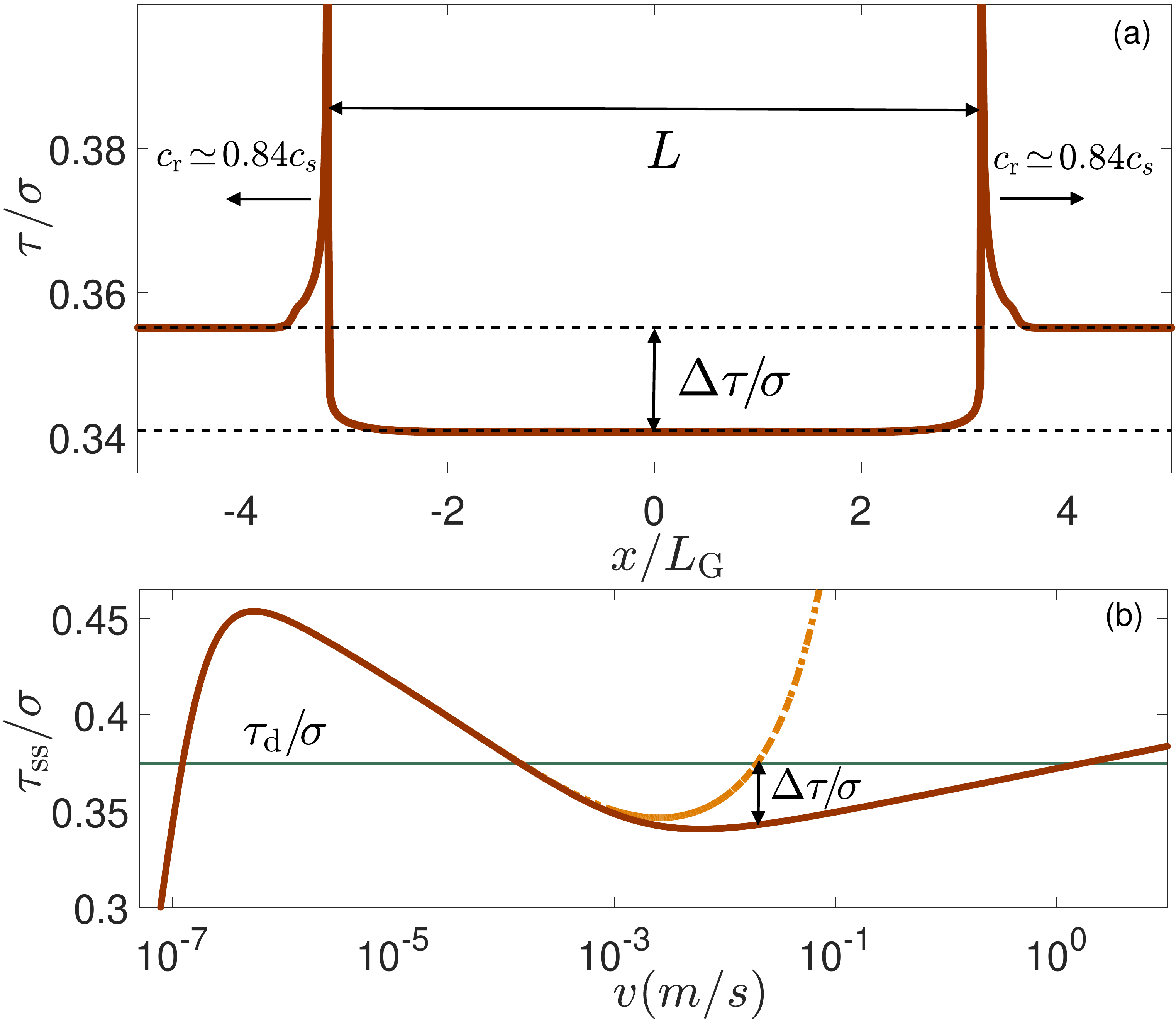}
  \caption{(a) A snapshot of the frictional stress $\tau(x)$ (normalized by the normal stress $\sigma$) during rupture propagation that emerges in dynamic simulations with the steady-state friction law shown in panel (b) and $\tau_{\rm d}\!=\!0.375\sigma$ (see text and~\citet{PartI} for additional details). The snapshot reveals two rapid rupture fronts (the rupture length $L$ is marked) propagating at an instantaneous speed $c_{\rm r}\!\simeq\!0.84c_s$ in opposite directions into regions characterized by the applied stress $\tau_{\rm d}$ and leaving behind them a well-defined residual stress $\tau_{\rm res}\!<\!\tau_{\rm d}$. Consequently, a well-defined and finite stress drop $\Delta\tau$ emerges, as marked. Note that the $y$-axis is truncated at $\tau/\sigma\!=\!0.4$ for visual clarity and that $x$ is normalized by a generalized Griffith-like length $L_{\rm G}$, defined in Eq.~\eqref{eq:c-L} (with a unity prefactor). (b) The steady-state friction stress $\tau_{\rm ss}(v)$, normalized by a constant normal stress $\sigma$, vs.~the slip rate $v$ (solid brown line). The curve has a generic $N$-shape~\citep{Bar-Sinai2014}, with a maximum at an extremely low $v$ and a minimum at an intermediate $v$. The horizontal line represents the driving stress $\tau_{\rm d}$, which intersects the $N$-shaped steady-state friction curve at three points; the leftmost and rightmost ones are stable fixed points, while the intermediate one is an unstable one. The effective steady-state friction curve (dash-dotted orange line) is obtained by adding $\frac{\mu}{2c_s}v$ (with $\mu\!=\!9$GPa and $c_s\!=\!2739$m/s) to the solid brown line, see~\citet{PartI} for more details. The stress drop $\Delta\tau$ of Eq.~\eqref{eq:SD}, which equals the one shown in panel (a), is marked by the black double-arrow.}\label{fig:Fig1}
\end{figure}

The theoretical prediction in Eq.~\eqref{eq:SD} has been supported by existing experimental results for rapid frictional rupture~\citep{PartI}, for times shorter than the waves reflection time from outer boundaries, and by computer simulations in infinite systems. An example taken from one of these computer simulations is presented in Fig.~\ref{fig:Fig1}a (cf.~Fig.~3 in~\citet{PartI}), where two rapid rupture fronts propagating in opposite directions are observed, leaving behind them a well-defined stress drop $\Delta\tau$ that quantitatively agrees with the theoretical predictions (see~\citet{PartI} for details). The most outstanding theoretical question that remains open in the context of the analogy between frictional rupture and ordinary cracks, once the necessary conditions associated with the emergence of a finite stress drop $\Delta\tau$ are met, is to what extent the analogy actually holds, both in qualitative and in quantitative terms. This question is systematically addressed in this paper.

The existence of a finite stress drop $\Delta\tau$ does not immediately guarantee that the analogy between frictional rupture and ordinary fracture holds because proper scale separation should also be satisfied. That is, the residual stress $\tau_{\rm res}$ behind the propagating rupture should be reached on a scale (typically termed the cohesive zone) much smaller than the rupture size $L$ (cf.~Fig.~\ref{fig:Fig1}a). If such scale separation is valid, we expect all crack-like properties to emerge in frictional rupture. In particular, we expect the frictional stress and slip velocity fields near the rupture edge to feature the famous square root singularity of conventional fracture mechanics~\citep{Freund1998}. Moreover, under these conditions, we expect the singularity-associated energy flux into the edge region to balance the edge-localized energy dissipation in excess of the power invested against the residual stress $\tau_{\rm res}$. This energy balance relation amounts to an effective equation of motion for rupture propagation~\citep{Freund1998}.

In this paper we show that for generic and realistic frictional constitutive relations, and once the conditions for the emergence of an effective crack-like behavior are met, frictional rupture dynamics are approximately --- yet quantitatively -- described by a crack-like, fracture mechanics energy balance equation~\citep{Freund1998}. This is achieved in a few steps. In Sect.~\ref{sec:length-velocity} we show that if one {\em assumes} the existence of the conventional square root singularity of ordinary fracture mechanics and the associated near-edge energy balance in frictional rupture, the latter follows a generic rupture length-velocity relation based on the knowledge of the stress drop $\Delta\tau$ alone. In Sect.~\ref{sec:EOM}, we quantitatively and systematically test these assumptions separately. We first show that the conventional square root singularity of standard fracture mechanics provides a good quantitative description of the near rupture edge stress and slip velocity fields simultaneously. We then propose a physically-motivated procedure to independently extract an effective fracture energy from the dissipative interfacial dynamics and show that it is balanced by the singularity-associated energy flux into the edge region to a good approximation.

These results indicate that the scale separation mentioned above is approximately satisfied for frictional rupture and that indeed the effective fracture energy corresponds to edge-localized dissipation. However, the proposed procedure to extract the relevant edge-localized dissipation allows us to show, also in Sect.~\ref{sec:EOM}, that there exists additional energy dissipation in excess of the power invested against the residual stress $\tau_{\rm res}$. This contribution to the energy dissipation associated with frictional rupture propagation is shown to be non-edge-localized, i.e.~to be spatially extended, and as such demonstrates interesting deviations from the ordinary crack-like framework. Finally, the significance and implications of our findings for various phenomena are briefly discussed in Sect.~\ref{sec:summary}. Together with the recent results about the emergence of stress drops in frictional rupture~\citep{PartI}, this work offers a comprehensive and basic understanding of why, how and to what extent frictional rupture might be viewed as an ordinary fracture process.

\section{Crack-like scaling and the dependence of the length-velocity relation on the stress drop}
\label{sec:length-velocity}

As explained above, and with the results of~\citet{PartI} in mind, we aim at carefully exploring the implications of stress drops --- once they exist --- for frictional dynamics. The expected implications, to be detailed below, directly follow from the analogy to ordinary fracture mechanics and consequently from its standard predictions~\citep{Freund1998,Svetlizky2019}. The challenge is to test whether these predictions are satisfied as emergent properties of the underlying physics without assuming them a priori. Some of these predictions have been previously studied in the literature~\citep{Cocco2002,Bizzarri2003,Das2003,Rubin2005,Chester2005,Tinti2005,Bizzarri2010a,Nielsen2016}, but to the best of our knowledge these studies have not yet led to a comprehensive picture of the analogy between frictional rupture and ordinary fracture.

The existence of a stress drop behind the two edges of propagating frictional rupture, cf.~Fig.~\ref{fig:Fig1}a, suggests that the load bearing capacity of the interface in this region is reduced, $\tau_{\rm res}\!<\!\tau_{\rm d}$, and consequently that parts of the interface ahead of the edges should compensate for this reduction, i.e.~carry stress that is larger than $\tau_{\rm d}$. In the framework of the classical theory of fracture, the so-called Linear Elastic Fracture Mechanics (LEFM), this stress amplification ahead of the rupture edges follows a universal singularity as the rupture edge is approached~\citep{Freund1998}
\begin{equation}
\label{eq:SIF}
\tau(x) \sim \frac{K(L,c_{\rm r})}{\sqrt{|x-x_{\rm r}|}}, \quad K(L,c_{\rm r}) \sim \Delta\tau\,\sqrt{L}\,{\cal K}(c_{\rm r}/c_s) \ ,
\end{equation}
where $K$ quantifies the intensity of the singularity (hence it is termed the stress intensity factor~\citep{Irwin1957}), $x_{\rm r}$ is the location of each of the rupture edges, $L$ is the instantaneous distance between the two edges (i.e.~the rupture length/size, cf.~Fig.~\ref{fig:Fig1}a) and ${\cal K}(c_{\rm r}/c_s)$ is a dimensionless function of the instantaneous propagation speed $c_{\rm r}$ of each edge. We note that here and below numerical pre-factors are omitted as we are interested in crack-like scaling relations in this section. In addition, the slip velocity is predicted to follow the very same singular behavior
\begin{equation}
v(x)\sim \frac{c_{\rm r}\,K(L,c_{\rm r})}{\mu\sqrt{|x-x_{\rm r}|}} \ ,
\label{eq:SIF1}
\end{equation}
just behind the edges (note the absolute value). As expected, the intensity of the amplification/singularity $K(L,c_{\rm r})$ in Eq.~\eqref{eq:SIF} increases with increasing $\Delta\tau$ and the rupture length $L$ ($L$ is the size of the region in which the interfacial load bearing capacity is reduced, hence a larger compensation/amplification exists). The relations in Eqs.~\eqref{eq:SIF}-\eqref{eq:SIF1} are valid independently of the symmetry mode of rupture, and in particular in the context of frictional rupture, they are valid for both in-plane shear (mode-II) and anti-plane shear (mode-III) symmetries.

Standard fracture mechanics predicts that the square root singularity in Eqs.~\eqref{eq:SIF}-\eqref{eq:SIF1} is accompanied by a finite flux of energy $G$ into the rupture edge region (known as the energy release rate~\citep{Irwin1957}, even though it is not a rate), taking the form~\citep{Irwin1957}
\begin{equation}
\label{eq:G}
G(L,c_{\rm r}) \sim {\cal A}(c_{\rm r}/c_s)\frac{\left[K(L,c_{\rm r})\right]^2}{\mu} \ ,
\end{equation}
where ${\cal A}(c_{\rm r}/c_s)$ is a known universal and dimensionless function that depends on the fracture symmetry mode (here mode-II or mode-III). Finally, by invoking energy balance in the edge region, standard fracture mechanics predicts that~\citep{Freund1998}
\begin{equation}
\label{eq:Gc}
G(L,c_{\rm r}) = G_{\rm c}(c_{\rm r}) \ ,
\end{equation}
where $G_{\rm c}(c_{\rm r})$ is the effective fracture energy (of dimensions of energy per unit area) associated with the transition from the $v\!\approx\!0$ state ahead of the edge to the $v\!>\!0$ state behind it, which possibly depends on the rupture speed $c_{\rm r}$. It is crucial to understand that unlike ordinary tensile (mode-I symmetry) fracture, where $G_{\rm c}(c_{\rm r})$ is the only dissipation in the problem, in the friction problem frictional dissipation exists {\em everywhere} along the sliding interface and not just in the transition region near the rupture edge. The way energy dissipation is partitioned in the friction problem will be discussed below.

The above discussion raises several basic questions; most notably, does the square root singularity of Eqs.~\eqref{eq:SIF}-\eqref{eq:SIF1} generically exist in frictional rupture once $\Delta\tau$ exists? Can the effective fracture energy $G_{\rm c}(c_{\rm r})$ be meaningfully separated from the entire dissipation associated with frictional motion? And if so, can the energy balance of Eq.~\eqref{eq:Gc} be verified by independently calculating both $G_{\rm c}$ and $G$ (the latter using Eq.~\eqref{eq:G})? While various aspects of these questions have certainly been addressed in the literature~\citep{Cocco2002,Bizzarri2003,Das2003,Rubin2005,Chester2005,Tinti2005,Bizzarri2010a,Nielsen2016}, we believe that systematically addressing all of them in a single system is still missing. Before performing such a systematic analysis, we address first a rather strong implication of the relations discussed above.
\begin{figure}[ht!]
  \centering
  \includegraphics[width=\columnwidth]{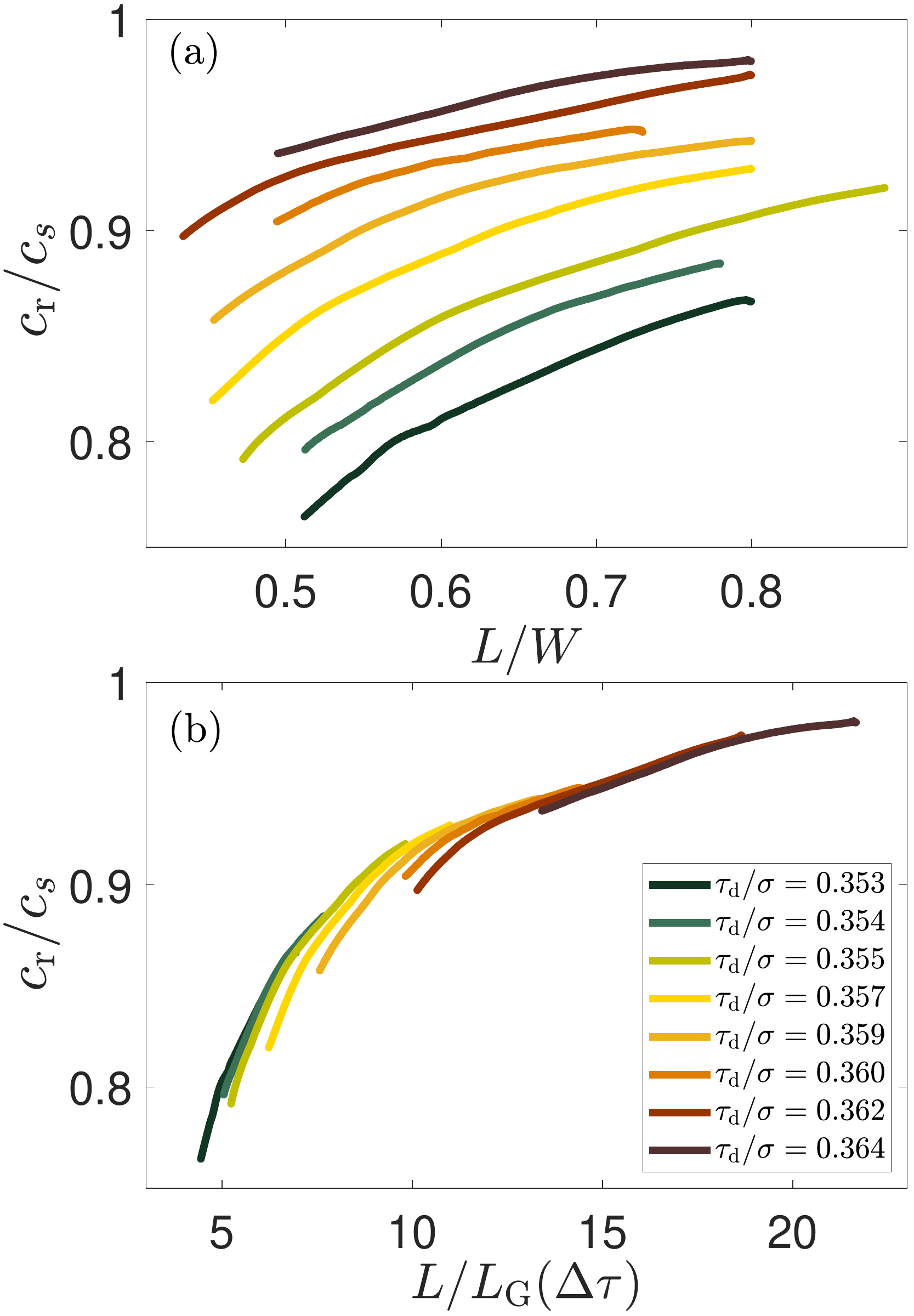}
  \caption{(a) The frictional rupture velocity $c_{\rm r}$, normalized by $c_s$, as a function of the frictional rupture length $L$ (normalized by the system size $W\!=\!320$m used in these calculations) for different driving stress levels $\tau_{\rm d}/\sigma$ (as detailed in the legend of panel (b)), using the $N$-shaped friction law of Fig.~\ref{fig:Fig1}b. Frictional rupture is nucleated as described in~\citet{PartI}. (b) The prediction of Eq.~\eqref{eq:c-L} is tested by plotting $c_{\rm r}/c_s$ vs.~$L/L_{\rm G}(\Delta\tau)$, where $\Delta\tau$ varies with $\tau_{\rm d}$ according to Eq.~\eqref{eq:SD} (see also Fig.~3c in~\citet{PartI}).
  $L_{\rm G}(\Delta\tau)$, as defined in Eq.~\eqref{eq:c-L}, is evaluated with $\mu\!=\!9$GPa, $G_{\rm c}\!=\!0.65$J/m$^2$ and a unity prefactor. The length-velocity curves of panel (a) all collapse on a master envelope curve as predicted by Eq.~\eqref{eq:c-L}, see additional discussion in the text.}\label{fig:Fig2}
\end{figure}

Combining Eqs.~\eqref{eq:SIF}-\eqref{eq:Gc}, one obtains the following stress drop dependent length-velocity relation
\begin{equation}
\label{eq:c-L}
c_{\rm r}/c_s = {\cal F}\left[L/L_{\rm G}(\Delta\tau)\right]\quad\hbox{with}\quad L_{\rm G}(\Delta\tau) \sim \frac{\mu\,G_{\rm c}}{\left(\Delta\tau\right)^2} \ ,
\end{equation}
which is valid under the {\em assumption} that $G_{\rm c}$ is independent of $c_{\rm r}$. Here $L_{\rm G}(\Delta\tau)$ is a generalized Griffith-like length~\citep{Andrews1976,Freund1998} and ${\cal F}(\cdot)$ is a monotonically increasing function that we do not specify.

To test this prediction, we employed the generic rate-and-state friction constitutive framework, presented in detail in~\citet{PartI}. Within this framework, the interfacial constitutive law at any position $x$ along the interface and at any time $t$ is described by the following local relation
\begin{equation}
\label{eq:friction_law}
\tau=\sigma\,\sgn(v)\,f(|v|,\phi) \ ,
\end{equation}
which must be supplemented with a dynamical equation for the evolution of $\phi$. Extensive evidence indicates that $\phi$ physically represents the age/maturity of the contact (hence it is related to the real contact area)~\citep{Rice1983,Marone1998,Nakatani2001,Baumberger2006,Dietrich2007,Nagata2012,Bhattacharya2014}, and that its evolution takes the form
\begin{equation}
\label{eq:dot_phi}
\dot\phi = g\left(\frac{|v|\phi}{D}\right) \ ,
\end{equation}
with $g(1)\=0$ and where $\phi$ is of time dimension. The characteristic slip displacement $D$ controls the transition from a stick state $v\!\approx\!0$, with a characteristic structural state $\phi\=\phi_0$, to a steadily slipping/sliding state $v\!>\!0$, with $\phi_{\rm ss}\=D/v$. The precise functional form of $g(\cdot)$ (with $g(1)\=0$) plays no role in what follows. The function $f(|v|,\phi_{\rm ss}\=D/v)\=\tau_{\rm ss}(v)/\sigma$, under steady-state sliding conditions and a controlled normal stress $\sigma$, has been measured over a broad range of slip rates $v$ for many materials~\citep{Baumberger2006}.

Together with general theoretical considerations~\citep{Bar-Sinai2014}, it is now established that the steady-state frictional stress $\tau_{\rm ss}(v)$ is generically $N$-shaped, as shown in Fig.~\ref{fig:Fig1}b (solid brown line). Finally, the effective friction curve obtained by adding the radiation damping term $\frac{\mu}{2c_s}v$, which has been shown to play an important role in the emergence of stress drops in frictional rupture~\citep{PartI}, is also presented in Fig.~\ref{fig:Fig1}b (dash-dotted orange line). We would like to stress that, as shown in~\citet{PartI}, pure velocity-weakening friction laws also effectively feature $N$-shaped behavior due to the radiation damping term (and hence also feature a finite stress drop). Consequently, the results to be presented below equally apply to velocity-weakening friction laws.

Coupling this constitutive framework to spectral boundary integral method~\citep{Geubelle1995,Morrissey1997,Breitenfeld1998} calculations in infinite systems under mode-III deformation conditions, gave rise to frictional rupture such as the one shown in~Fig.~\ref{fig:Fig1}a. In this approach, the displacement field ${\bm u}(x,y,t)\=u_z(x,y,t)\hat{\bm z}$ (the unit vectors satisfy $\hat{\bm z}\,\bot\,\hat{\bm x},\hat{\bm y}$) is computed at the interface $y\!\to\!0^\pm$ self-consistently with the far-field stress $\tau_{\rm d}$ and the friction law of Eq.~\eqref{eq:friction_law}, see~\citet{PartI} for additional details. Based on such numerical computations, we plot in Fig.~\ref{fig:Fig2}a the normalized frictional rupture velocity $c_{\rm r}/c_s$ vs.~the frictional rupture length $L$ for various driving stress levels $\tau_{\rm d}$ (detailed in the legend of Fig.~\ref{fig:Fig2}b). The different $c_{\rm r}(L)$ curves span a rather broad range. Equation~\eqref{eq:c-L} predicts that these curves can be collapsed onto a master curve if $L$ is rescaled by $L_{\rm G}(\Delta\tau)$, where $\Delta\tau(\tau_{\rm d})$ is given in Eq.~\eqref{eq:SD} (see also Fig.~3c in~\citet{PartI}) and the effective fracture energy $G_{\rm c}$ is assumed to be independent of $c_{\rm r}$. To follow this rescaling procedure, $L_{\rm G}(\Delta\tau)$ of Eq.~\eqref{eq:c-L} is evaluated with a unity prefactor, $\mu\!=\!9$GPa and $G_{\rm c}\!=\!0.65$J/m$^2$. The way to extract the value of the effective fracture energy $G_{\rm c}$ is discussed in Sect.~\ref{sec:EOM} below. The outcome of the rescaling procedure is presented in Fig.~\ref{fig:Fig2}b.

It is observed that the different $c_{\rm r}(L)$ curves, which exhibited a rather large spread in Fig.~\ref{fig:Fig2}a, collapse on the envelope of a single master curve upon rescaling $L$ by $L_{\rm G}(\Delta\tau)$. Note that deviations from the master curve are observed at early times (small $L$ values in each curve); this is expected as the crack-like behavior cannot be valid in the nucleation stage, but rather only when $L$ is sufficiently large and frictional rupture is sufficiently well-developed. The collapse in Fig.~\ref{fig:Fig2}b provides indirect, yet strong, support to the applicability of the crack-like relations in Eqs.~\eqref{eq:SIF}-\eqref{eq:Gc} to frictional rupture. These relations will be directly tested next.

\section{The emergence of stress singularity and local energy balance}
\label{sec:EOM}

\begin{figure*}[ht!]
  \includegraphics[width=\textwidth]{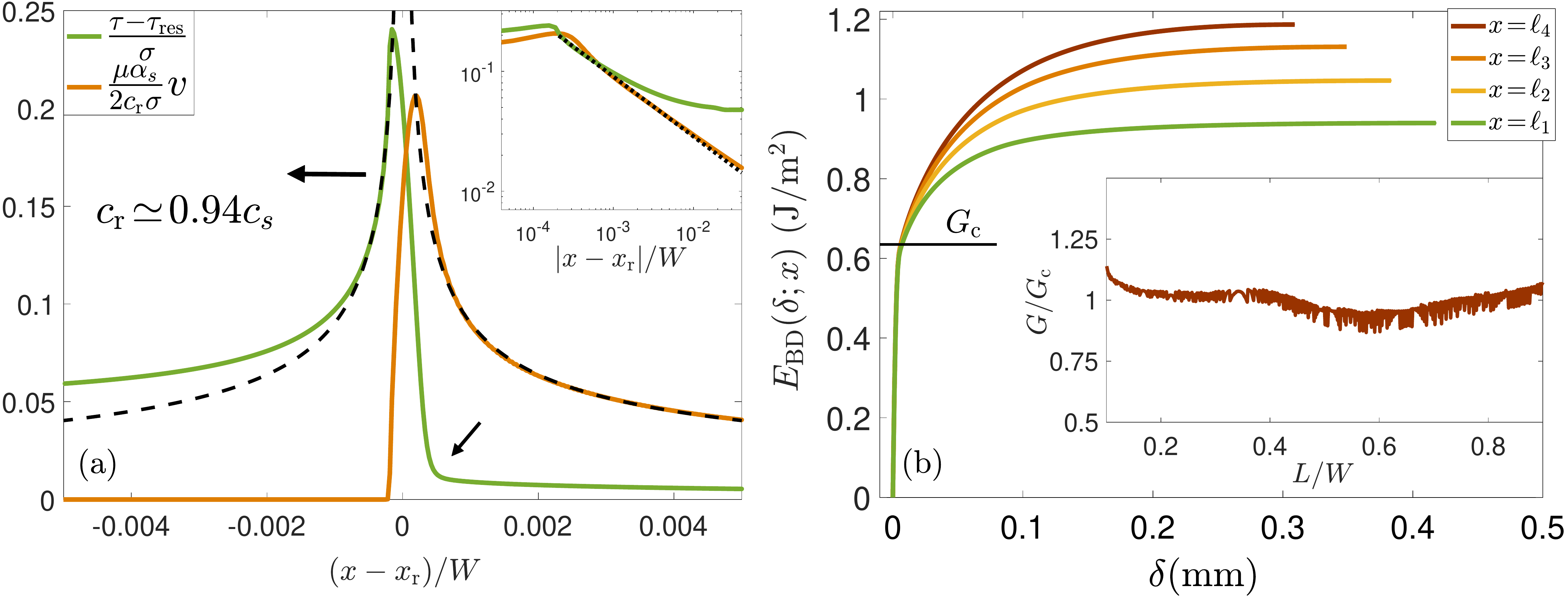}
  \caption{(a) The normalized spatial profiles of $\tau(x,t)$ and $v(x,t)$ near a rupture edge propagating from right to left with a velocity $c_{\rm r}\!\simeq\!0.94c_s$ at time $t$. $x$ is shifted by $x_{\rm r}$, which corresponds to the location of effective rupture edge (cf.~Eqs.~\eqref{eq:SIF}-\eqref{eq:SIF1}). Both fields are normalized/shifted by quantities defined in the text, except for $\alpha_s\!\equiv\!\sqrt{1-c_{\rm r}^2/c_s^2}$. The dashed lines are the results of fitting the solid lines to Eqs.~\eqref{eq:SIF}-\eqref{eq:SIF1}, with $K\!=\!64$kPa$\cdot$m$^{1/2}$, see~\citet{SM} for additional details. The tilted arrow is discussed in the text. (inset) The same as the main panel, but on a double logarithmic scale with the $x$-axis being $|x-x_{\rm r}|$. Note that since the dashed lines in the main panel are symmetric with respect to $x_{\rm r}$, using $|x-x_{\rm r}|$ implies the existence of a single dashed line in the inset. The inset highlights both the quality of the fit and the different spatial ranges used for each field, see~\citet{SM} for additional details. (b) The breakdown energy $E_{\rm BD}(\delta; x)$, defined in Eq.~\eqref{eq:E_BD}, vs.~slip $\delta$ for $4$ interfacial locations $x\!=\!\ell_i$, with $\ell_1/W\!=\!0.15$, $\ell_2/W\!=\!0.20$, $\ell_3/W\!=\!0.25$ and $\ell_4/W\!=\!0.30$. $\ell_i$ are measured from the nucleation site (the center of the system) and the system size is $W\!=\!80$m. The horizontal black line marks the splitting of the different curves, which is identified with $G_{\rm c}\!\approx\!0.65$J/m$^2$. (inset) $G/G_{\rm c}$ vs.~$L/W$, where $L$ is the rupture length. $G$ is calculated using $K(L)$, cf.~panel (a) and the~\citet{SM}, through Eq.~\eqref{eq:G} and $G_{\rm c}$ is calculated through Eq.~\eqref{eq:Gc_integral}. The generic properties of the results presented in this figure are independent of the details of the friction law (not shown).}\label{fig:Fig3}
\end{figure*}

One of the major implications of the existence of a finite stress drop $\Delta\tau$ is the emergence of stress singularity near the frictional rupture edge, as explained above and as formulated in Eqs.~\eqref{eq:SIF}-\eqref{eq:SIF1}. In order to directly test this prediction, we present in Fig.~\ref{fig:Fig3}a the (properly normalized) spatial profiles of $\tau(x,t)$ and $v(x,t)$ near a rupture edge at time $t$. We then fit the two fields {\em together} to Eqs.~\eqref{eq:SIF}-\eqref{eq:SIF1}, demanding the {\em same} stress intensity factor $K$ and the {\em same} effective tip location $x_{\rm r}$ (the details of the fitting procedure are extensively discussed in the~\citet{SM}).

The resulting fits are superimposed on the fields $\tau(x,t)$ and $v(x,t)$ in Fig.~\ref{fig:Fig3}a. The square root singular behavior faithfully describes the two fields near the front edge, supporting the prediction that such a singular behavior emerges in the presence of a finite stress drop $\Delta\tau$. Note that the spatial range in which the fields are described by the square root singular behavior is larger for the slip velocity $v(x,t)$ than for the frictional stress $\tau(x,t)$. The reason is that $\tau(x,t)$ features a significantly narrower range of values between its peak value and the applied stress $\tau_{\rm d}$ (in the large $|x|$ limit) compared to the corresponding range for $v(x,t)$, and thus the latter can accommodate a singular behavior, which is by construction an intermediate asymptotic behavior, over a larger spatial range.

The results of Fig.~\ref{fig:Fig3}a demonstrate that a rather well-defined stress intensity factor $K(L,c_{\rm r})$ is associated with frictional rupture in the presence of a finite stress drop $\Delta\tau$, from which the energy release rate $G(L,c_{\rm r})$ can be readily extracted using Eq.~\eqref{eq:G}~\citep{SM}. Next, in order to test the validity of Eq.~\eqref{eq:Gc}, we need to independently calculate the effective fracture energy $G_{\rm c}$ associated with frictional rupture propagation. To this aim, we define the energy per unit area that is dissipated at a given interfacial location $x$ during the transition from a non-slipping/sticking state to a steadily sliding state characterized by the residual stress $\tau_{\rm res}$~\citep{Bizzarri2010a}
\begin{equation}
E_{\rm BD}(\delta; x)=\int_0^\delta \big(\tau(\delta')-\tau_{\rm res}\big)d\delta' \ .
\label{eq:E_BD}
\end{equation}
Here the slip history at a location $x$ is given by the slip displacement $\delta(x,t)\!\equiv\!u_z(x,y\=0^+,t)\!-\!u_z(x,y\=0^-,t)$, where $\dot\delta(x,t)\=v(x,t)$, and the subscript 'BD' stands for 'breakdown'. The breakdown energy quantifies the excess dissipation on top of the frictional dissipation associated with sliding against the residual stress $\tau_{\rm res}$. Note that we cannot a priori identify the breakdown energy defined in Eq.~\eqref{eq:E_BD} with the effective fracture energy $G_{\rm c}$, as will be discussed next.

In Fig.~\ref{fig:Fig3}b we plot the breakdown energy $E_{\rm BD}(\delta; x)$ at $4$ different interfacial locations $x\=\ell_i$, $i\=1\!-\!4$, ordered by their proximity to the nucleation site (the center of the domain). It is observed that $E_{\rm BD}(\delta; x)$ perfectly overlaps for the different locations $x$'s at small $\delta$, but exhibits location dependence at significantly larger $\delta$, where it levels off to different limiting values that become closer to one another as $x$ increases. These observations can be understood as follows; the frictional stress $\tau(x,t)$ presented in Fig.~\ref{fig:Fig3}a exhibits two distinct behaviors behind the propagating rupture edge (here the propagation is from right to left). First, it features a strong decay well within the edge region. Second, as denoted by the arrow, there exists a transition to a slow decay towards $\tau_{\rm res}$ on a significantly larger lengthscale, extending far beyond the edge region (the full spatial extent of this decay is not shown). This slow spatial decay stems from the rate and state dependence of the friction law, which implies that all of the interfacial fields in the problem $\tau(x,t), v(x,t), \phi(x,t)$ slowly approach their respective asymptotic steady-state values $\tau_{\rm res}, v_{\rm res}, D/v_{\rm res}$. Finally, as rupture propagation in the presence of a finite stress drop is intrinsically out of steady state, i.e.~rupture accelerates towards $c_s$ as shown in Fig.~\ref{fig:Fig2}, we expect some position dependence of $E_{\rm BD}(\delta; x)$. This dependence should become weaker as the limiting velocity $c_{\rm r}\!\to\!c_s$ is approached, as is indeed observed in Fig.~\ref{fig:Fig3}b.

The physical picture emerging from the above discussion suggests that the location independent part of the breakdown energy $E_{\rm BD}(\delta; x)$, which is associated with excess dissipation near the rupture edge, should be identified as the effective fracture energy $G_{\rm c}$ appearing in Eq.~\eqref{eq:Gc}. This idea is pictorially demonstrated by the horizontal black line in Fig.~\ref{fig:Fig3}b, which identifies $G_{\rm c}$ with the point in which the various $E_{\rm BD}(\delta; x)$ curves start to split/deviate one from another (from which a value of $G_{\rm c}\!\approx\!0.65$J/m$^2$ can be inferred). To make the identification of $G_{\rm c}$ more quantitative and to allow a direct test of Eq.~\eqref{eq:Gc}, we invoke the observation that the combination $v\phi/D$ strongly overshoots unity in the edge region ($v\phi/D\!>\!1$ implies $\dot\phi\!<\!0$, which is associated with contact area reduction), then slightly undershoots it and finally approaches unity from below far from the edge~\citep{SM}. We note that the position of the first crossing $v\phi/D\=1$ approximately corresponds to the position marked by small arrow in Fig.~\ref{fig:Fig3}a. Consequently, the edge-localized dissipation $G_{\rm c}$ can be estimated as the excess dissipation associated with the spatial region for which $v\phi/D\!>\!1$, quantified by the following spatial integral
\begin{equation}
G_{\rm c}(c_{\rm r}) \equiv \frac{1}{c_{\rm r}(t)} \int_{v\phi/\!D>1} \!\big(\tau(x,t)-\tau_{\rm res} \big)\,v(x,t)\,dx \ .
\label{eq:Gc_integral}
\end{equation}
We note that this estimate of $G_{\rm c}$ appears to be consistent with an analytic approximation available in the literature~\citep{Cocco2002, Bizzarri2003, Rubin2005}, which may shed light on the dependence of $G_{\rm c}$ on interfacial parameters (see~\citet{SM} for details).

We are now in a position to directly test Eq.~\eqref{eq:Gc}, where the energy release rate $G$ is calculated using the stress intensity factor extracted as shown in Fig.~\ref{fig:Fig3}a and $G_{\rm c}$ through Eq.~\eqref{eq:Gc_integral}. In the inset of Fig.~\ref{fig:Fig3}b, we plot the ratio $G/G_{\rm c}$ as a function of the rupture length $L$. It is observed that $G/G_{\rm c}$ is close to unity throughout the rupture propagation history, lending strong support to the ideas developed above. In particular, it shows that the rupture edge energy balance in Eq.~\eqref{eq:Gc} provides quantitative approximations for frictional rupture dynamics.

At the same time, our results also clearly demonstrate that $E_{\rm BD}(\delta; x)$ can be quite significantly larger than $G_{\rm c}$ and position dependent, implying that non-edge-localized dissipation in excess of the power invested against the residual stress $\tau_{\rm res}$ is a generic property of frictional interfaces featuring rate and state dependent friction. A similar physical situation has been discussed in~\citet{Brener2002}. That is, while a physically sensible extraction of the edge-localized excess dissipation $G_{\rm c}$ allows to obtain reasonably well quantitative approximations for frictional rupture dynamics based on the analogy to ordinary fracture, our results clearly indicate that this analogy is incomplete and that interesting deviations exist. These deviations are intimately related to the spatially extended (non-edge-localized) rate and state dependence of frictional interfaces, an intrinsic frictional property that is entirely absent in ordinary fracture, and are manifested in non-edge-localized excess dissipation. The latter may have important implications for the energy budget associated with frictional dynamics, and might be relevant to geophysical observations and their interpretations~\citep{Das2003,Chester2005,Tinti2005,Nielsen2016}.

\section{Summary and concluding remarks}
\label{sec:summary}

In this paper we set out to further explore the analogy between frictional rupture and ordinary fracture. The starting point for this investigation is our own very recent work that elucidated the physical origin of stress drops $\Delta\tau$ in frictional rupture~\citep{PartI}, which constitute a necessary condition for the analogy. Our major goal was to understand to what extent the analogy holds, both in qualitative and in quantitative terms, for interfaces described by generic and realistic frictional constitutive relations, once stress drops do exist.

We showed that for rate and state constitutive relations, frictional rupture dynamics are approximately --- yet quantitatively --- described by an ordinary fracture energy balance equation, when the conditions for the emergence of a finite stress drop $\Delta\tau$ are satisfied. To establish the quantitative status of this fracture mechanics energy balance equation, we proposed a physical criterion for extracting the rupture edge-localized dissipation directly from the frictional dynamics, allowing to define an effective fracture energy $G_{\rm c}$ for frictional problems. Surprisingly, we discovered that $G_{\rm c}$ does not account for all of the energy dissipation $E_{\rm BD}$ in excess of the energy dissipated against the residual stress $\tau_{\rm res}$ (cf.~Eq.~\eqref{eq:E_BD}). These findings imply that the analogy between frictional rupture and ordinary fracture is not complete, as manifested by the existence of a non-edge-localized contribution to $E_{\rm BD}$.

The difference between $E_{\rm BD}$ and $G_{\rm c}$ is intimately related to the generic rate and state dependence of friction, which is responsible for the two-step nature of the stress relaxation/weakening process associated with frictional rupture propagation; first, there exists a rather sharp stress drop that takes place over a relatively small slip, bringing the stress close to, but not identically to, the residual stress $\tau_{\rm res}$. Second, there exists a slower, longer-term process that brings the stress to the residual stress $\tau_{\rm res}$ over significantly larger slip. The latter stress relaxation/weakening process, which some authors attribute to melting or thermal pressurization~\citep{Rice2006,Viesca2015} not taken into account in the present work, is responsible for the difference between $E_{\rm BD}$ and $G_{\rm c}$. This physical picture is reminiscent of the model proposed in~\citet{kanamori2000}, and further discussed in~\citet{Abercrombie2005}, in trying to resolve some puzzling observations in relation to the energy budget of earthquake rupture. Moreover, this physical picture is consistent with~\citet{Chester2005} and~\citet{Tinti2005}, which concluded based on seismic data that the breakdown energy can be larger than the fracture energy for large earthquake ruptures. These results offer insight into open questions concerning earthquake energy budget~\citep{Das2003,Abercrombie2005,Chester2005,Tinti2005,Nielsen2016} and deserve additional investigation.

More generally, we expect our results to provide a conceptual and quantitative framework to address various fundamental and applied problems in relation to the rupture dynamics of frictional interfaces, with implications for both laboratory and geophysical-scale phenomena. For example, our results and theoretical framework are expected to apply also to slip pulses. Indeed, recent preliminary results, see Fig.~S6 in~\citet{Brener2018}, support this expectation.\\

{\em Acknowledgements} E.~B.~and J.-F.M.~acknowledge support from the
Rothschild Caesarea Foundation. E.~B.~acknowledges support
from the Israel Science Foundation (Grant No.~295/16). J.-F.M., F.~B.~and T.~R.~acknowledge support from the Swiss
National Science Foundation (Grant No.~162569). This research is made possible in part by the historic generosity of the Harold Perlman Family.

\clearpage

\onecolumngrid
\begin{center}
	\textbf{\large Supplemental Material for: ``The emergence of crack-like behavior of frictional rupture: Edge singularity and energy balance''}
\end{center}

\setcounter{equation}{0}
\setcounter{figure}{0}
\setcounter{section}{0}
\setcounter{table}{0}
\setcounter{page}{1}
\makeatletter
\renewcommand{\theequation}{S\arabic{equation}}
\renewcommand{\thefigure}{S\arabic{figure}}
\renewcommand{\thesubsection}{S-\arabic{subsection}}
\renewcommand{\thesection}{S-\arabic{section}}
\renewcommand*{\thepage}{S\arabic{page}}
\renewcommand{\bibnumfmt}[1]{[S#1]}
\renewcommand{\citenumfont}[1]{S#1}
\twocolumngrid

The goal of this document is to provide additional technical details regarding the extraction of the near-edge singular fields (Fig.~3a in the manuscript) and the effective fracture energy $G_{\rm c}$ from the interfacial dynamics (Fig.~3b in the manuscript), both discussed in Sect.~III of the manuscript. This is achieved in two steps; first, in Sect.~\ref{sec:equ_mot}, some relevant concepts and methodology are being discussed and tested using a conventional cohesive zone model of ordinary fracture. Then, in Sect.~\ref{sec:rupt_rands}, these concepts and tools are generalized for frictional rupture along interfaces described by generic friction constitutive relations, and additional details about their application in Sect.~III of the manuscript are briefly provided. The numerical tools and the generic interfacial constitutive relation (including the material parameters) are presented in~\cite{PartI,Brener2018}.

\section{Edge singularity and energy balance in a conventional cohesive zone model of ordinary fracture}
\label{sec:equ_mot}

Our goal here is to first develop the procedure for extracting the near-edge singular fields in a simpler case, where there is no residual stress (i.e.~ordinary fracture), where the Linear Elastic Fracture Mechanics (LEFM) singularity is regularized on a small lengthscale (i.e.~proper scale separation is realized) and the fracture energy $G_{\rm c}$ is prescribed. This is achieved by the well-known framework of cohesive zone crack models, attributed to Dugdale~\cite{Dugdale1960} and Barenblatt~\cite{Barenblatt1962}, which became very popular in the numerical modeling of dynamic
fracture (see, for example,~\cite{Breitenfeld1998,Barras2014}). Within this framework, we employ a linear slip-weakening cohesive law in which the strength of the interface $\tau^{\rm str}$ linearly reduces to zero over a characteristic slip displacement $\delta_c$
\begin{equation}
\tau^{\rm str}(x,t) = \tau_c\;\{1-\delta(x,t)/\delta_c\} \ ,
\label{equ:coh_law}
\end{equation}
where $\tau_{c}$ is the failure strength (determining the rupture peak stress), $\delta(x,t)$ is the slip displacement, and $\{ \xi \} \= \xi$
if $\xi\!>\!0$ and $0$ otherwise ($\xi$ is a dummy variable used to define the function $\{\cdot\}$ in Eq.~\eqref{equ:coh_law}). The linear slip-weakening law of
Eq.~\eqref{equ:coh_law} corresponds to a prescribed value of the fracture energy
\begin{equation}
 G_{\rm c} = \int_0^{\delta_c} \tau d\delta=\frac{1}{2}\tau_c\delta_c \ .
 \label{equ:gc_coh}
\end{equation}

\begin{figure}[ht]
\centering
\includegraphics[width=\linewidth]{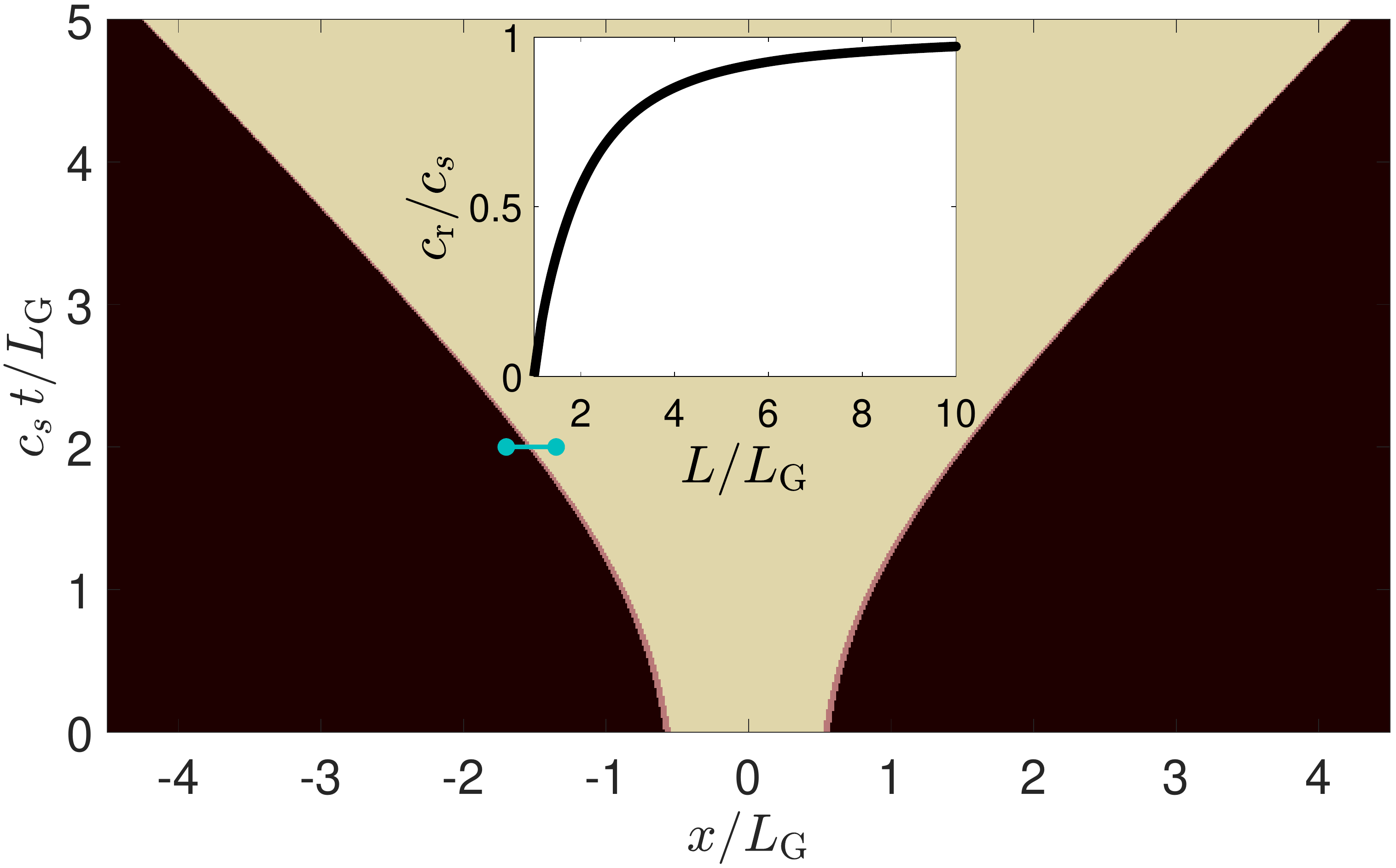}
\caption{Space-time diagram of the dynamic mode-III rupture event described in the text. The yellow region corresponds to the broken interface left behind the propagating rupture edges, the narrow red region corresponds to the cohesive zone and the black region corresponds to the intact interface. The blue line marks
the instant at which the snapshots of the stress and slip velocity fields in Fig.~\ref{fig:energy_balance}a are taken. (inset) The time evolution of the
rupture speed $c_{\rm r}$ as function of its size $L$.}
\label{fig:st_diagram}
\end{figure}
The spectral boundary integral method under mode-III symmetry (where the basic object is the out-of-plane displacement field at the interface, $u_z(x,y\=0,t)$, see manuscript and references therein for details) can be coupled to
Eq.~\eqref{equ:coh_law} (i.e.~the latter replaces the friction law used in the manuscript) to generate propagating rupture fronts. In
this context, rupture is nucleated at the center of an interface at rest under a uniform shear
stress $\tau_{\rm d}$, where $0\!<\!\tau_{\rm d}\!<\!\tau_c$, by progressively increasing an originally infinitesimal
seed crack toward a critical size $L\=L_G$. The latter, known as the
\textit{Griffith critical length}~\cite{Andrews1976,Freund1998}, is given by (see also Eq.~(6) in the manuscript)
\begin{equation}
 L_G = \frac{4\mu\,G_{\rm c}}{\pi\,\tau^2_{\rm d}} \ ,
\label{eq:Griffith1}
\end{equation}
for mode-III cracks. In Fig.~\ref{fig:st_diagram}, we present the resulting dynamics that feature a crack that progressively accelerates toward $c_s$, the maximal admissible
rupture speed for mode-III symmetry.

\begin{figure*}[ht]
\centering
\includegraphics[width=\linewidth]{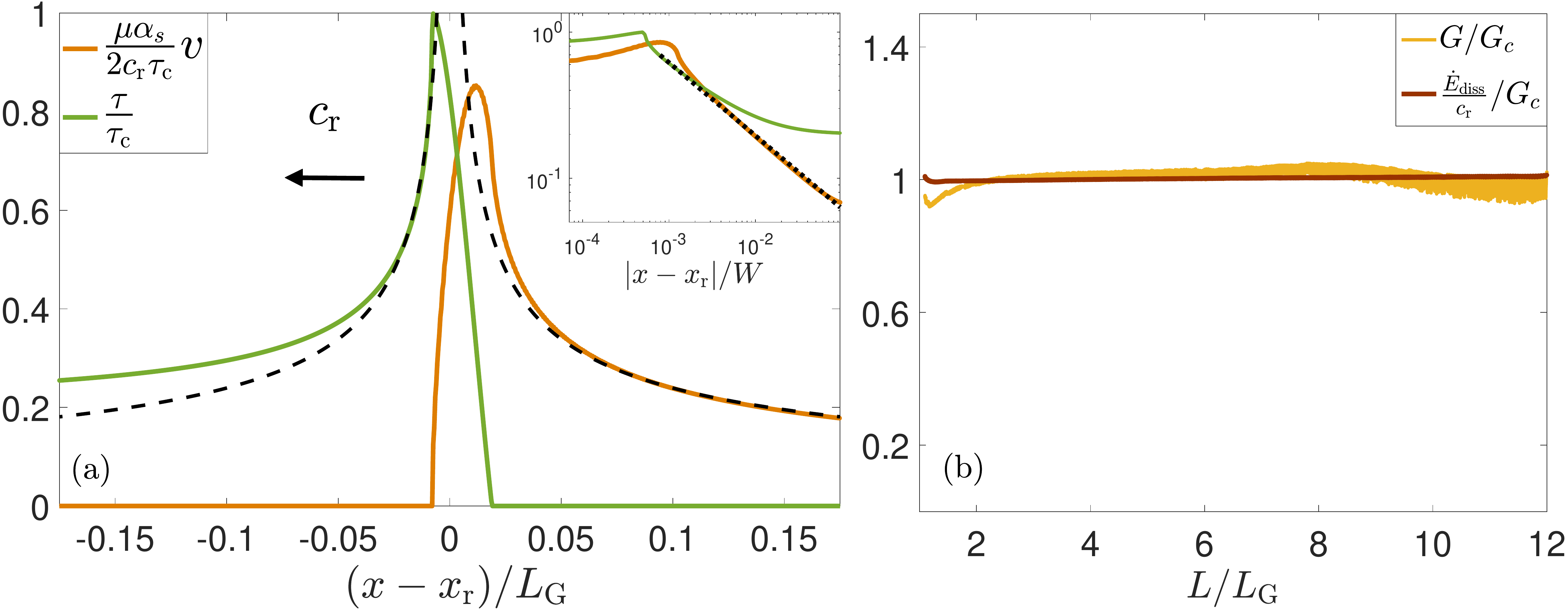}
\caption{(a) A snapshot of the normalized stress and slip velocity fields (see legend and the left-hand-sides of Eqs.~\eqref{equ:tau_sing}-\eqref{equ:v_sing}) near the edge of a rupture propagating at a speed $c_{\rm r}$ to the left (the snapshot corresponds to the blue horizontal line in Fig.~\ref{fig:st_diagram}, where rupture propagation in the simple slip-weakening cohesive zone model is presented). Note that $\tau_{\rm c}$ is used to nondimensionalize the fields and that $\tau_{\rm res}\!=\!0$ in this case. The black dashed lines correspond to fits to Eqs.~\eqref{equ:tau_sing}-\eqref{equ:v_sing}, see text for additional details. (inset) The same as the main panel, but on a double logarithmic scale and the $x$-axis is $|x-x_{\rm r}|/W$, see text for additional details. (b) $G$ and $\dot{E}_{\textmd{diss}}/c_{\rm r}$, both normalized by $G_{\rm c}$, are plotted as a function of the normalized rupture size $L/L_{\rm G}$ (see legend in order to distinguish the different curves). These quantities are discussed in detail in the text.}
\label{fig:energy_balance}
\end{figure*}

The instantaneous rate of dissipated energy associated with the propagation of one rupture edge (recall that there are two of these) can be obtained as~\cite{Barras2014}
\begin{equation}
\dot{E}_{\mathrm{diss}}(t)=\int_0^{\frac{1}{2}\!W}\!\tau(x,t)\,v(x,t) \,dx \ ,
\label{equ:dotE}
\end{equation}
where $W$ is the system size. The integral attains a finite contribution only inside the well-defined cohesive zone near the propagating rupture edge, where both $\tau(x,t)$ and $v(x,t)$ are non-zero.
The cohesive zone (also termed \textit{fracture process zone} in ordinary fracture), which corresponds to the region where the stress $\tau(x,t)$ drops from the peak stress (failure strength) $\tau_c$ to $0$, is
marked by the red-shaded region in Fig.~\ref{fig:energy_balance}a. A snapshot of the stress $\tau(x,t)$ and slip velocity $v(x,t)$ distributions near the propagating rupture edge are also presented in Fig.~\ref{fig:energy_balance}a (and see also Fig.~\ref{fig:st_diagram}). The fracture energy, defined in Eq.~\eqref{equ:gc_coh}, is the energy
dissipated per unit crack extension $dL$
\begin{equation}
G_{\rm c}(t) = \frac{d}{d L}E_{\mathrm{diss}}(t)=\frac{dE_{\mathrm{diss}}}{d t}\Big/\frac{dL}{dt} =\frac{\dot{E}_{\mathrm{diss}}(t)}{c_{\rm r}(t)} \ ,
\label{equ:edr}
\end{equation}
which is constant for the slip-weakening model used here (see Fig.~\ref{fig:energy_balance}b).

Standard fracture theory predicts that close to the propagating rupture edges, we have the famous square root singular fields~\cite{Freund1998}
\begin{equation}
\tau(r\=x_{\rm r}\!-\!x,\theta\=0,c_{\rm r})-\tau_{\rm res} \simeq
\frac{K_{\rm III}}{\sqrt{2\pi (x_{\rm r}-x)}}
\label{equ:tau_sing}
\end{equation}
and
\begin{equation}
\frac{\mu\,\alpha_s(c_{\rm r})}{2c_{\rm r}}v(r\=x\!-\!x_{\rm r},\theta\=\pi,c_{\rm r}) \simeq
\frac{K_{\rm III}}{\sqrt{2\pi (x-x_{\rm r})}} \ ,
\label{equ:v_sing}
\end{equation}
where $(r,\theta)$ is a polar coordinate system moving with the rupture edge, $\alpha_s(c_{\rm r})\=\sqrt{1-c_{\rm r}^2/c_s^2}$, $x_{\rm r}$ is the effective edge location and $K_{\rm III}$ is the mode-III stress intensity factor. We subtracted the residual stress $\tau_{\rm res}$ from the frictional stress field such that the shifted stress field vanishes behind the rupture edge and normalized the slip velocity field such that the left-hand-sides of both Eqs.~\eqref{equ:tau_sing}-\eqref{equ:v_sing} attain comparable values; note that for the slip-weakening model used here we have $\tau_{\rm res}\=0$, and it makes no difference, but in general one may have $\tau_{\rm res}\!>\!0$ (also in the framework of slip-weakening models), see Sect.~\ref{sec:rupt_rands}. In addition, we used $v\=2\dot{u}_z$ since $v$ is the slip velocity, not the particle (mass) velocity $\dot{u}_z$. Finally, as is evident from the right-hand-sides of both Eqs.~\eqref{equ:tau_sing}-\eqref{equ:v_sing}, the normalized slip velocity $v$ and frictional stress $\tau$ fields are symmetric functions relative to $x_{\rm r}$ (i.e.~it is the very same function of $|x-x_{\rm r}|$), though the spatial ranges in which the singular form is valid differ for the two fields. This issue will be discussed below, where we explain how the two free parameters in Eqs.~\eqref{equ:tau_sing}-\eqref{equ:v_sing} --- $x_{\rm r}$ and $K_{\rm III}$ --- are determined. We stress that the proper normalization and shift used in Eqs.~\eqref{equ:tau_sing}-\eqref{equ:v_sing} allow us to consider the stress and slip velocity fields on equal footing.

The square root singularity is associated with a finite energy flux into the edge region, the so-called energy release rate $G$, which for mode-III symmetry takes the form~\cite{Freund1998}
\begin{equation}
 G(t) = \frac{1}{\alpha_s}\frac{K_{\rm III}^2}{2\mu} \ .
 \label{equ:err}
\end{equation}
Our goal now is to extract the stress intensity factor from the singular fields of Eqs.~\eqref{equ:tau_sing}-\eqref{equ:v_sing}, to use Eq.~\eqref{equ:err} to calculate $G$ and to check whether the near-edge energy balance $G\=G_{\rm c}$ is satisfied. As all of the assumptions of conventional fracture theory are satisfied by the model, the energy balance equation should be satisfied.

We start by estimating the stress intensity factor from the near-edge stress and slip velocity distributions shown in Fig.~\ref{fig:energy_balance}a. That is, we fit the normalized and shifted near-edge stress and slip velocity fields to the singular form in Eqs.~\eqref{equ:tau_sing}-\eqref{equ:v_sing}, with $x_{\rm r}$ and $K_{\rm III}$ as the two free parameters. To make the procedure well defined, we also need to specify the spatial range over which the fits are performed. In determining the spatial range of the fit of the two fields, several physical considerations are invoked; first, it is clear that the fits cannot include the regions where the fields (cf.~the examples in Fig.~\ref{fig:energy_balance}a) attain their peak values as these are associated with the regularization of the singular behavior (the cohesive zone). Second, the fitting ranges cannot extend too far away from the edge region as the fields there include also non-singular contributions. Finally, as the overall variability of the stress field is smaller compared to that of the slip velocity field, we expect the singular region to be narrower for the former. We employ a nonlinear least-squares regression fitting procedure~\cite{Jones2001} to determine the best estimates for $x_{\rm r}$ and $K_{\rm III}$, and selected the fitting ranges to be as large as possible within the constraints imposed by the physical considerations just stated.

The resulting fits, i.e.~the right-hand-sides of Eqs.~\eqref{equ:tau_sing}-\eqref{equ:v_sing}, are superimposed on the normalized slip velocity $v$ and frictional stress $\tau$ fields in Fig.~\ref{fig:energy_balance}a (dashed lines). To highlight the spatial fitting ranges used, we replot the results in Fig.~\ref{fig:energy_balance}a on a double logarithmic scale against $|x-x_{\rm r}|/W$ in the inset (note that due to the symmetry of the singular form on the right-hand-sides of Eqs.~\eqref{equ:tau_sing}-\eqref{equ:v_sing}, we have now a single fit that describes the two fields over different spatial ranges). The inset shows that the spatial fitting ranges for the two fields are different, that the range for the slip velocity field is wider than the one for the frictional stress field and that the peak regions are properly excluded. Finally, we verified that the values of $x_{\rm r}$ and $K_{\rm III}$ are robust against changes in the spatial fitting ranges {\em within} the stated constraints.

The extracted value of $K_{\rm III}$ has been used to calculate the energy release rate $G$ according to Eq.~\eqref{equ:err}. Then we applied the fitting procedure to the whole rupture propagation history and the a priori known value of $G_{\rm c}$ in Eq.~\eqref{equ:gc_coh} has been used to plot in Fig.~\ref{fig:energy_balance}b $G/G_{\rm c}$ as a function of $L/L_{\rm G}$, where $L$ is the rupture length. The results strongly support the expected relation $G/G_{\rm c}\=1$ and hence also validate our fitting procedure. Note that some deviation from $G/G_{\rm c}\=1$ is observed, reflecting some uncertainly in the singular behavior, even in simple slip-weakening models. Finally, for completeness, we also plot in Fig.~\ref{fig:energy_balance}b $\dot{E}_{\mathrm{diss}}(t)/c_{\rm r}(t)$ of Eq.~\eqref{equ:edr}, normalized by $G_{\rm c}$, which indeed equals unity throughout the rupture propagation process, as expected. The same fitting procedure is applied in the manuscript to the frictional rupture dynamics of interfaces described by rate-and-state friction, as discussed next.

\section{Application to the frictional rupture dynamics of interfaces described by rate-and-state friction}
\label{sec:rupt_rands}

A procedure similar to the one described in the previous section is applied in the manuscript to the frictional rupture dynamics of interfaces described by rate-and-state friction. However, the differences between the simple slip-weakening cohesive zone model discussed in the previous section and the more realistic rate-and-state friction models discussed in the manuscript, which are intimately related to the central question addressed in the manuscript, call for some modifications that will be discussed here. First, frictional rupture features a finite residual stress $\tau_{\rm res}\!>\!0$ under some conditions (extensively discussed in~\cite{PartI}). That is, the strength of the interface does not drop to zero behind the rupture front as in the simple slip-weakening cohesive zone model (note that in general slip-weakening cohesive zone models can definitely feature a constant residual stress $\tau_{\rm res}$), but rather attains a finite value (on what lengthscale this value is attained is yet another central question addressed in the manuscript). The linearity of the elastodynamic field equations~\cite{Palmer1973} implies that the driving stress $\tau_{\rm d}$ in the ordinary fracture case should be simply replaced by the stress drop $\Delta\tau\=\tau_{\rm d}-\tau_{\rm res}\!>\!0$ in the frictional case. This implies that $\tau_{\rm res}$ should be subtracted from the stress field $\tau(x,t)$ before fitting it to the square root singular contribution in Eq.~\eqref{equ:tau_sing} (cf.~Fig.~3a in the manuscript). Moreover, this implies that a generalization of the Griffith length of Eq.~\eqref{eq:Griffith1} takes the form
\begin{equation}
 L_G = \frac{4\mu\,G_{\rm c}}{\pi\left(\Delta\tau\right)^2} \ ,
\label{eq:Griffith2}
\end{equation}
which is identical to the corresponding expression in Eq.~(6) in the manuscript, up to the dimensionless and order unity pre-factor $4/\pi$.

As discussed in the manuscript, the generalized Griffith-like length in Eq.~\eqref{eq:Griffith2} and in Eq.~(6) in the manuscript highlights another difference between simple slip-weakening cohesive zone models and rate-and-state friction models related to $G_{\rm c}$. While in slip-weakening cohesive zone models $G_{\rm c}$ is an a priori prescribed quantity, in rate-and-state friction models the existence and identification of a well-defined $G_{\rm c}$ from the interfacial dynamics is not obvious. That is, one should understand whether and how an effective fracture energy $G_{\rm c}$ can be properly defined, and what the associated lengthscale is. A procedure to define and extract $G_{\rm c}$ is discussed and employed in the manuscript. Here we supplement it with additional rationalization and details.
\begin{figure}[ht]
\centering
\includegraphics[width=\columnwidth]{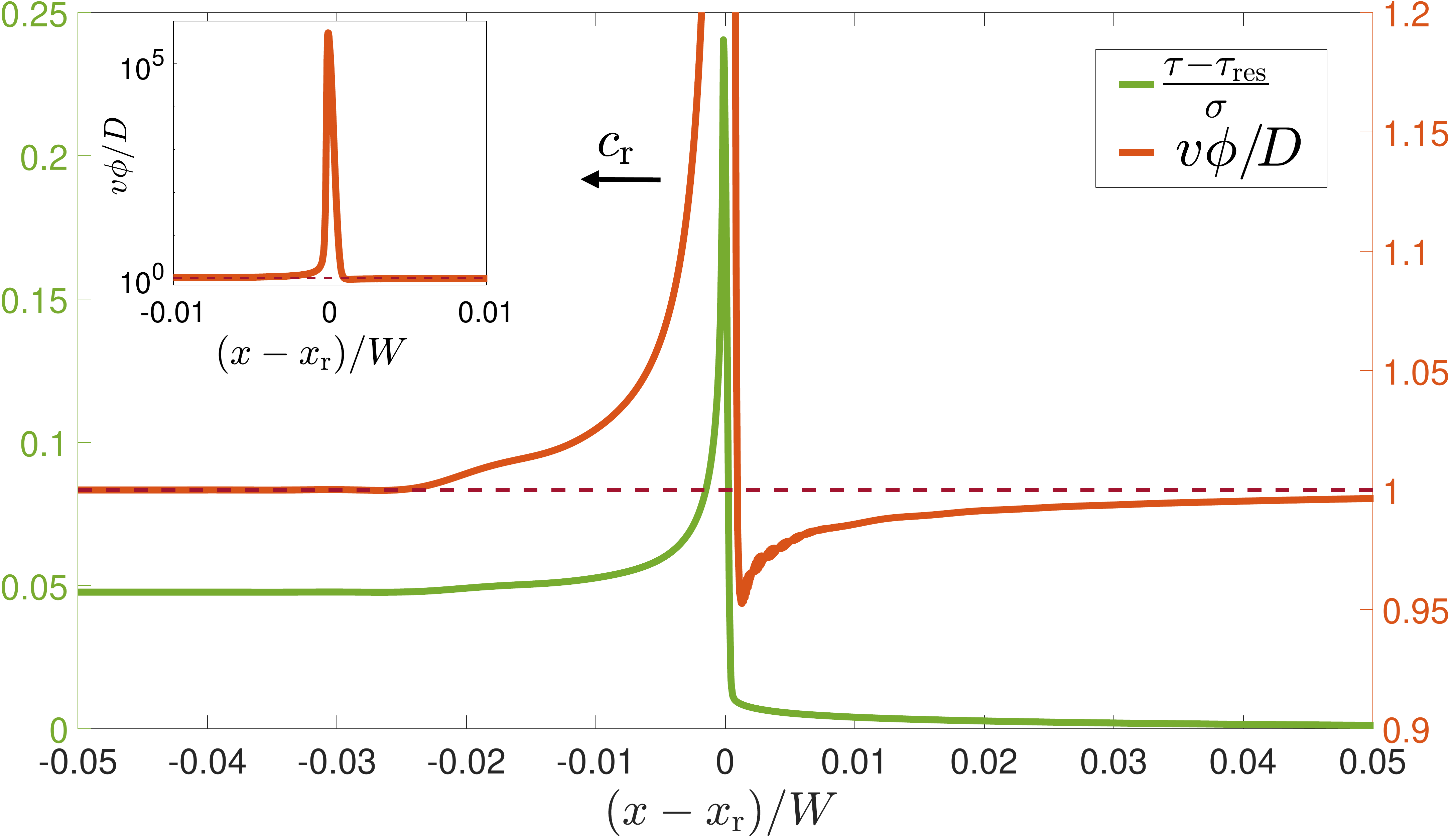}
\caption{A snapshot of the properly normalized (see legend) stress field $\tau(x,t)$ (left $y$-axis) and $v(x,t)\phi(x,t)/D$ (right $y$-axis) corresponding to the solution presented in Fig.~3a in the
manuscript, where the $y$-axis is truncated to allow the properties of the fields near the rupture edge to be visible. (inset) $v(x,t)\phi(x,t)/D$ near the rupture edge without truncating the $y$-axis.}
\label{fig:vphiD}
\end{figure}

The basic idea is related to the observation that the frictional stress $\tau(x,t)$ follows two distinct relaxation regimes in the wake of rupture fronts, as demonstrated in Fig.~3a in the manuscript. It first undergoes a rather strong initial drop that is followed by a slow decay towards $\tau_{\rm res}$. Such behavior is inherent to the rate-and-state dependence of the frictional strength~\cite{Baumberger2006}. The initial strong drop is associated with a rather localized region near the rupture edge (see arrow in Fig.~3a in the manuscript) and the slow
decay towards $\tau_{\rm res}$ is characterized by a much larger lengthscale. We consequently proposed that the former should be associated with the effective fracture energy $G_{\rm c}$.

\begin{figure}[ht]
\centering
\includegraphics[width=\columnwidth]{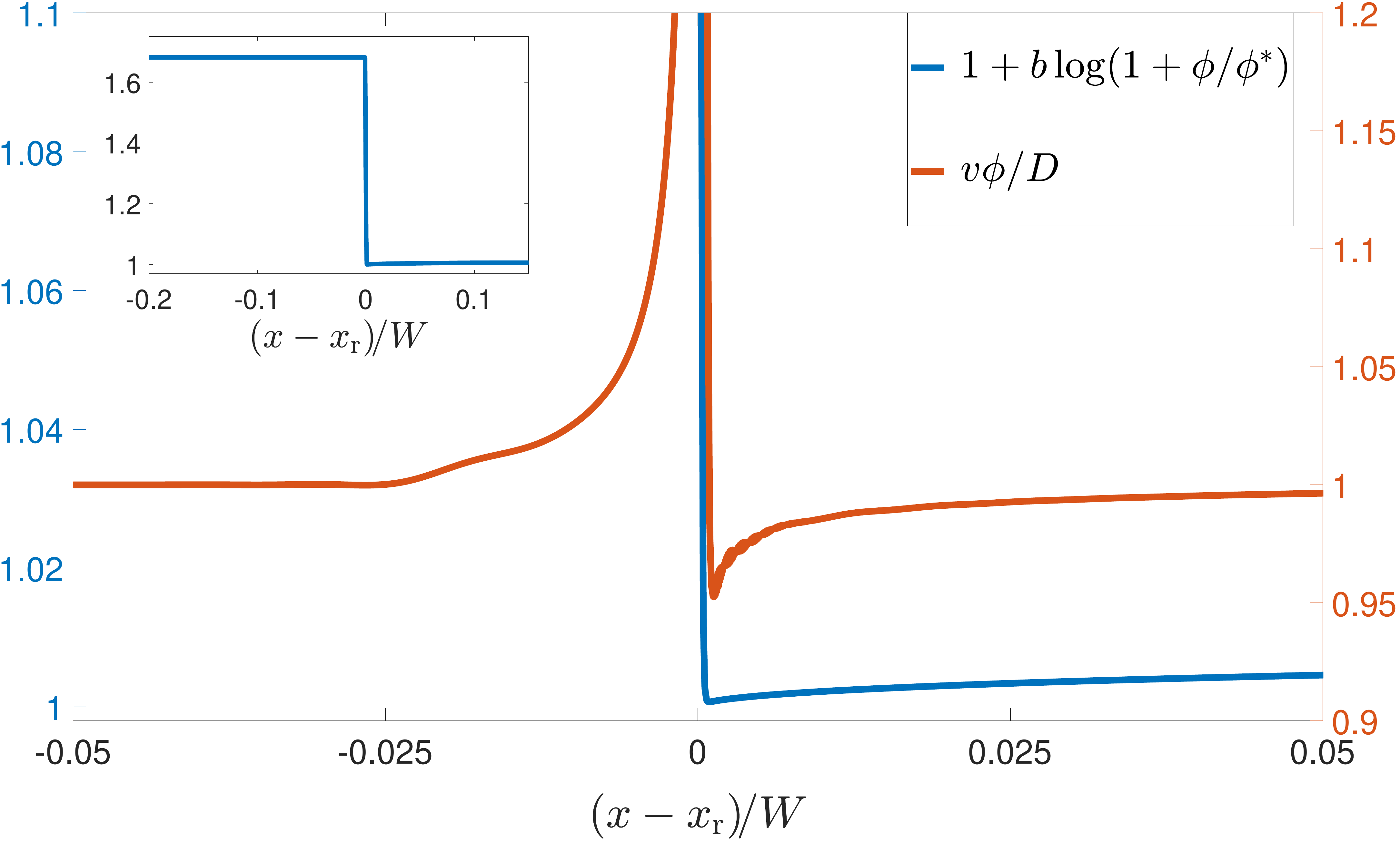}
\caption{A snapshot of the real contact area $A_{\rm r}(x,t)\!\sim\!1+b\log[1+\phi(x,t)/\phi^*]$ (blue line, left $y$-axis) corresponding to $v(x,t)\phi(x,t)/D$ of Fig.~\ref{fig:vphiD}, which is reproduced here (orange line, right $y$-axis). The real contact area also exhibits slow relaxation to its asymptotic value behind the rupture edge.
(inset) A full scale plot of $A_{\rm r}(x,t)\!\sim\!1+b\log[1+\phi(x,t)/\phi^*]$ near the rupture edge, directly demonstrating that the latter is associated with a reduction of the real contact area.}
\label{fig:contact_area}
\end{figure}

In order to formalize this idea and to make the extraction of $G_{\rm c}$ quantitative, we focus on the dimensionless combination $v(x,t)\phi(x,t)/D$, which is shown in Fig.~\ref{fig:vphiD} and which according to Eq.~(8) in the manuscript controls the evolution of the structural state of the interface $\phi(x,t)$. The latter is known to determine the real contact area $A_{\rm r}(x,t)\!\sim\!1+b\log[1+\phi(x,t)/\phi^*]$ of the interface~\cite{Baumberger1999} (for the definition of the parameters $b$ and $\phi^*$, and their values used here, see~\cite{PartI,Brener2018}). Hence, it is directly related to the rupture process, involving a transition from an initial value of $A_{\rm r}$ ahead of the rupture front to a significantly lower value behind it (see the inset of Fig.~\ref{fig:contact_area}). This transition corresponds to a transition between $v\phi/D\=1$ ahead of the rupture front, with a very small $v$ and hence a large $\phi$, and $v\phi/D\=1$ behind it, with a large $v$ and hence a much smaller $\phi$. In between, $v\phi/D$ is expected to attain significantly larger values. This physical picture is demonstrated in the inset of Fig.~\ref{fig:vphiD}, which corresponds to the rupture front shown in Fig.~3a in the manuscript.

The two-step nature of the approach of $v\phi/D$ to its steady-state is revealed in the main panel of Fig.~\ref{fig:vphiD}, which presents a zoomed in version of the inset. The figure reveals that after the huge peak in $v\phi/D$, which occurs on a small lengthscale near the rupture edge, $v\phi/D$ undershoots unity and then approaches unity slowly from below, on a significantly larger lengthscale. We consequently attribute the small lengthscale weakening process to the near-edge dissipation $G_{\rm c}$, i.e.~to the effective fracture energy, where the additional dissipation associated with the larger lengthscale is discussed in the manuscript. In quantitative terms, this picture implies that $G_{\rm c}$ is estimated through the dissipation corresponding to $v(x,t)\phi(x,t)/D\!>\!1$, as formulated in Eq.~(10) in the manuscript.

The latter criterion is demonstrated in Fig.~\ref{fig:vphiD}, where the frictional stress $\tau(x,t)$ of Fig.~3a in the manuscript is superimposed on $v(x,t)\phi(x,t)/D$, to exactly correspond to the change in the relaxation behavior of $\tau(x,t)$ towards $\tau_{\rm res}$ that was discussed above. This criterion is also in line with recent physics-based interpretations of rate-and-state friction formulations~\cite{Baumberger1999,Bar-Sinai2014,Molinari2019}. Finally, for completeness, we present in Fig.~\ref{fig:contact_area} a snapshot of the spatial distribution of the real contact area $A_{\rm r}(x,t)\!\sim\!1+b\log[1+\phi(x,t)/\phi^*]$~\cite{Baumberger2006}.

We note that the estimation of $G_{\rm c}$ through the dissipation corresponding to the criterion $v(x,t)\phi(x,t)/D\!>\!1$ appears to be consistent with available analytic approximations for the effective fracture energy~\cite{Cocco2002,Bizzarri2003,Rubin2005}. In particular, the expression
\begin{equation}
  G_{\rm c} = \frac{D\sigma}{2}\frac{\partial f(|v|,\phi)}{\partial\log\left(\phi\right)} \left[\log(v_{\rm c}/v_{\rm bg})\right]^2
\label{eq:RA2005}
\end{equation}
has been proposed in~\cite{Rubin2005}. Here $\partial f(|v|,\phi)/\partial\log\left(\phi\right)$ is the aging coefficient ($f(|v|,\phi)$ is the friction law introduced in Eq.~(7) in the manuscript), $v_{\rm bg}$ corresponds to the steady-state velocity in the stick state (prior to the arrival of the rupture front) and $v_{\rm c}$ is the slip velocity far behind the rupture front. We estimate $v_{\rm bg}$ as the leftmost intersection point in Fig.~1b in the manuscript, i.e.~$v_{\rm bg}\!\approx\!10^{-7}$m/s, and $v_{\rm c}$ as the rightmost intersection point with the effective steady-state friction curve, i.e.~$v_{\rm c}\!\approx\!10^{-2}$m/s. Using the parameters used in this work (see~\cite{Brener2018}), i.e.~$D\=0.5\!\times\!10^{-6}$m, $\sigma\=10^6$Pa and $\partial f(|v|,\phi)/\partial\log\left(\phi\right)\=0.021$ (the latter equals $b f_0$ in the notation of~\cite{Brener2018}), and plugging everything in Eq.~\eqref{eq:RA2005}, we obtain $G_{\rm c}\!\approx\!0.7$J/m$^2$. The latter is in reasonably good agreement with $G_{\rm c}$ of Fig.~3b in the manuscript. In order to further substantiate this agreement, future work should extend the comparison by systematically varying the parameters involved.

To conclude, the procedure to extract the singular contribution of near-edge fields and to test the energy balance relation $G\=G_{\rm c}$ presented in Sect.~\ref{sec:equ_mot} is applied in the manuscript to rate-and-state frictional interfaces. In this case, $\tau_{\rm d}$ is replaced by the stress drop $\Delta\tau$ and $G_{\rm c}$ is estimated from the interfacial dynamics according to Eq.~(10) in the manuscript, as explained in detail here.


\begin{thebibliography}{47}
\expandafter\ifx\csname natexlab\endcsname\relax\def\natexlab#1{#1}\fi
\providecommand{\url}[1]{\texttt{#1}}
\providecommand{\href}[2]{#2}
\providecommand{\path}[1]{#1}
\providecommand{\DOIprefix}{doi:}
\providecommand{\ArXivprefix}{arXiv:}
\providecommand{\URLprefix}{URL: }
\providecommand{\Pubmedprefix}{pmid:}
\providecommand{\doi}[1]{\href{http://dx.doi.org/#1}{\path{#1}}}
\providecommand{\Pubmed}[1]{\href{pmid:#1}{\path{#1}}}
\providecommand{\bibinfo}[2]{#2}
\ifx\xfnm\undefined \def\xfnm[#1]{\unskip,\space#1}\fi
\bibitem[{SM()}]{SM}
\bibinfo{title}{{See Supplemental Material for additional information}}, .
\bibitem[{Abercrombie and Rice(2005)}]{Abercrombie2005}
\bibinfo{author}{Abercrombie\xfnm[ R.E.]}, \bibinfo{author}{Rice\xfnm[ J.R.]}.
\newblock \bibinfo{title}{{Can observations of earthquake scaling constrain
  slip weakening?}}
\newblock \bibinfo{journal}{Geophys J Int}
  \bibinfo{year}{2005};\bibinfo{volume}{162}(\bibinfo{number}{2}):\bibinfo{pages}{406--424}.
\newblock \URLprefix
  \url{https://academic.oup.com/gji/article-lookup/doi/10.1111/j.1365-246X.2005.02579.x}.
  \DOIprefix\doi{10.1111/j.1365-246X.2005.02579.x}.
\bibitem[{Andrews(1976)}]{Andrews1976}
\bibinfo{author}{Andrews\xfnm[ D.J.]}.
\newblock \bibinfo{title}{{Rupture propagation with finite stress in antiplane
  strain}}.
\newblock \bibinfo{journal}{J Geophys Res}
  \bibinfo{year}{1976};\bibinfo{volume}{81}(\bibinfo{number}{20}):\bibinfo{pages}{3575--3582}.
\newblock \URLprefix \url{http://doi.wiley.com/10.1029/JB081i020p03575}.
  \DOIprefix\doi{10.1029/JB081i020p03575}.
\bibitem[{Bar-Sinai et~al.(2014)Bar-Sinai, Spatschek, Brener and
  Bouchbinder}]{Bar-Sinai2014}
\bibinfo{author}{Bar-Sinai\xfnm[ Y.]}, \bibinfo{author}{Spatschek\xfnm[ R.]},
  \bibinfo{author}{Brener\xfnm[ E.A.]}, \bibinfo{author}{Bouchbinder\xfnm[
  E.]}.
\newblock \bibinfo{title}{{On the velocity-strengthening behavior of dry
  friction}}.
\newblock \bibinfo{journal}{J Geophys Res Solid Earth}
  \bibinfo{year}{2014};\bibinfo{volume}{119}(\bibinfo{number}{3}):\bibinfo{pages}{1738--1748}.
\newblock \URLprefix \url{http://doi.wiley.com/10.1002/2013JB010586}.
  \DOIprefix\doi{10.1002/2013JB010586}.
\bibitem[{Barras et~al.(2019)Barras, Aldam, Roch, Brener, Bouchbinder and
  Molinari}]{PartI}
\bibinfo{author}{Barras\xfnm[ F.]}, \bibinfo{author}{Aldam\xfnm[ M.]},
  \bibinfo{author}{Roch\xfnm[ T.]}, \bibinfo{author}{Brener\xfnm[ E.A.]},
  \bibinfo{author}{Bouchbinder\xfnm[ E.]}, \bibinfo{author}{Molinari\xfnm[
  J.F.]}.
\newblock \bibinfo{title}{{The emergence of crack-like behavior of frictional
  rupture: The origin of stress drops}}. \newblock \bibinfo{journal}{To appear in Physical Review X} \bibinfo{year}{2019}. \newblock \URLprefix
  \url{http://arxiv.org/abs/1906.11533}.
  \href{http://arxiv.org/abs/1906.11533}{\tt arXiv:1906.11533}.
\bibitem[{Baumberger and Caroli(2006)}]{Baumberger2006}
\bibinfo{author}{Baumberger\xfnm[ T.]}, \bibinfo{author}{Caroli\xfnm[ C.]}.
\newblock \bibinfo{title}{{Solid friction from stick–slip down to pinning and
  aging}}.
\newblock \bibinfo{journal}{Adv Phys}
  \bibinfo{year}{2006};\bibinfo{volume}{55}(\bibinfo{number}{3-4}):\bibinfo{pages}{279--348}.
\newblock \URLprefix
  \url{http://www.tandfonline.com/doi/abs/10.1080/00018730600732186}.
  \DOIprefix\doi{10.1080/00018730600732186}.
\bibitem[{Bayart et~al.(2015)Bayart, Svetlizky and Fineberg}]{Bayart2015}
\bibinfo{author}{Bayart\xfnm[ E.]}, \bibinfo{author}{Svetlizky\xfnm[ I.]},
  \bibinfo{author}{Fineberg\xfnm[ J.]}.
\newblock \bibinfo{title}{{Fracture mechanics determine the lengths of
  interface ruptures that mediate frictional motion}}.
\newblock \bibinfo{journal}{Nat Phys}
  \bibinfo{year}{2015};\bibinfo{volume}{12}(\bibinfo{number}{2}):\bibinfo{pages}{166--170}.
\newblock \URLprefix \url{https://www.nature.com/articles/nphys3539}.
  \DOIprefix\doi{10.1038/nphys3539}.
\bibitem[{Ben-Zion(2008)}]{Ben-Zion2008}
\bibinfo{author}{Ben-Zion\xfnm[ Y.]}.
\newblock \bibinfo{title}{{Collective behavior of earthquakes and faults:
  Continuum-discrete transitions, progressive evolutionary changes, and
  different dynamic regimes}}.
\newblock \bibinfo{journal}{Rev Geophys}
  \bibinfo{year}{2008};\bibinfo{volume}{46}(\bibinfo{number}{4}):\bibinfo{pages}{RG4006}.
\newblock \URLprefix \url{http://doi.wiley.com/10.1029/2008RG000260}.
  \DOIprefix\doi{10.1029/2008RG000260}.
\bibitem[{Bhattacharya and Rubin(2014)}]{Bhattacharya2014}
\bibinfo{author}{Bhattacharya\xfnm[ P.]}, \bibinfo{author}{Rubin\xfnm[ A.M.]}.
\newblock \bibinfo{title}{{Frictional response to velocity steps and 1-D fault
  nucleation under a state evolution law with stressing-rate dependence}}.
\newblock \bibinfo{journal}{J Geophys Res Solid Earth}
  \bibinfo{year}{2014};\bibinfo{volume}{119}(\bibinfo{number}{3}):\bibinfo{pages}{2272--2304}.
\newblock \URLprefix \url{http://doi.wiley.com/10.1002/2013JB010671}.
  \DOIprefix\doi{10.1002/2013JB010671}.
\bibitem[{Bizzarri(2010)}]{Bizzarri2010a}
\bibinfo{author}{Bizzarri\xfnm[ A.]}.
\newblock \bibinfo{title}{{On the relations between fracture energy and
  physical observables in dynamic earthquake models}}.
\newblock \bibinfo{journal}{J Geophys Res}
  \bibinfo{year}{2010};\bibinfo{volume}{115}(\bibinfo{number}{B10}):\bibinfo{pages}{B10307}.
\newblock \URLprefix \url{http://doi.wiley.com/10.1029/2009JB007027}.
  \DOIprefix\doi{10.1029/2009JB007027}.
\bibitem[{Bizzarri and Cocco(2003)}]{Bizzarri2003}
\bibinfo{author}{Bizzarri\xfnm[ A.]}, \bibinfo{author}{Cocco\xfnm[ M.]}.
\newblock \bibinfo{title}{{Slip-weakening behavior during the propagation of
  dynamic ruptures obeying rate- and state-dependent friction laws}}.
\newblock \bibinfo{journal}{J Geophys Res}
  \bibinfo{year}{2003};\bibinfo{volume}{108}(\bibinfo{number}{B8}):\bibinfo{pages}{2373}.
\newblock \URLprefix \url{http://doi.wiley.com/10.1029/2002JB002198}.
  \DOIprefix\doi{10.1029/2002JB002198}.
\bibitem[{Bizzarri and Liu(2016)}]{Bizzarri2016}
\bibinfo{author}{Bizzarri\xfnm[ A.]}, \bibinfo{author}{Liu\xfnm[ C.]}.
\newblock \bibinfo{title}{{Near-field radiated wave field may help to
  understand the style of the supershear transition of dynamic ruptures}}.
\newblock \bibinfo{journal}{Phys Earth Planet Inter}
  \bibinfo{year}{2016};\bibinfo{volume}{261}:\bibinfo{pages}{133--140}.
\newblock \URLprefix
  \url{https://linkinghub.elsevier.com/retrieve/pii/S0031920116300838}.
  \DOIprefix\doi{10.1016/j.pepi.2016.05.013}.
\bibitem[{Breitenfeld and Geubelle(1998)}]{Breitenfeld1998}
\bibinfo{author}{Breitenfeld\xfnm[ M.S.]}, \bibinfo{author}{Geubelle\xfnm[
  P.H.]}.
\newblock \bibinfo{title}{{Numerical analysis of dynamic debonding under 2D
  in-plane and 3D loading}}.
\newblock \bibinfo{journal}{Int J Fract}
  \bibinfo{year}{1998};\bibinfo{volume}{93}(\bibinfo{number}{1/4}):\bibinfo{pages}{13--38}.
\newblock \URLprefix \url{http://link.springer.com/10.1023/A:1007535703095}.
  \DOIprefix\doi{10.1023/A:1007535703095}.
\bibitem[{Brener et~al.(2018)Brener, Aldam, Barras, Molinari and
  Bouchbinder}]{Brener2018}
\bibinfo{author}{Brener\xfnm[ E.A.]}, \bibinfo{author}{Aldam\xfnm[ M.]},
  \bibinfo{author}{Barras\xfnm[ F.]}, \bibinfo{author}{Molinari\xfnm[ J.F.]},
  \bibinfo{author}{Bouchbinder\xfnm[ E.]}.
\newblock \bibinfo{title}{{Unstable Slip Pulses and Earthquake Nucleation as a
  Nonequilibrium First-Order Phase Transition}}.
\newblock \bibinfo{journal}{Phys Rev Lett}
  \bibinfo{year}{2018};\bibinfo{volume}{121}(\bibinfo{number}{23}):\bibinfo{pages}{234302}.
\newblock \URLprefix
  \url{https://link.aps.org/doi/10.1103/PhysRevLett.121.234302}.
  \DOIprefix\doi{10.1103/PhysRevLett.121.234302}.
\bibitem[{Brener and Marchenko(2002)}]{Brener2002}
\bibinfo{author}{Brener\xfnm[ E.A.]}, \bibinfo{author}{Marchenko\xfnm[ V.I.]}.
\newblock \bibinfo{title}{{Frictional shear cracks}}.
\newblock \bibinfo{journal}{J Exp Theor Phys Lett}
  \bibinfo{year}{2002};\bibinfo{volume}{76}(\bibinfo{number}{4}):\bibinfo{pages}{211--214}.
\newblock \URLprefix \url{http://link.springer.com/10.1134/1.1517386}.
  \DOIprefix\doi{10.1134/1.1517386}.
\bibitem[{Casado(2017)}]{Casado2017}
\bibinfo{author}{Casado\xfnm[ S.]}.
\newblock \bibinfo{title}{{Studying friction while playing the violin:
  exploring the stick–slip phenomenon}}.
\newblock \bibinfo{journal}{Beilstein J Nanotechnol}
  \bibinfo{year}{2017};\bibinfo{volume}{8}:\bibinfo{pages}{159--166}.
\newblock \URLprefix
  \url{http://www.beilstein-journals.org/bjnano/content/8/1/16}.
  \DOIprefix\doi{10.3762/bjnano.8.16}.
\bibitem[{Chester et~al.(2005)Chester, Chester and Kronenberg}]{Chester2005}
\bibinfo{author}{Chester\xfnm[ J.S.]}, \bibinfo{author}{Chester\xfnm[ F.M.]},
  \bibinfo{author}{Kronenberg\xfnm[ A.K.]}.
\newblock \bibinfo{title}{{Fracture surface energy of the Punchbowl fault, San
  Andreas system}}.
\newblock \bibinfo{journal}{Nature}
  \bibinfo{year}{2005};\bibinfo{volume}{437}(\bibinfo{number}{7055}):\bibinfo{pages}{133--136}.
\newblock \URLprefix \url{http://www.nature.com/articles/nature03942}.
  \DOIprefix\doi{10.1038/nature03942}.
\bibitem[{Cocco and Bizzarri(2002)}]{Cocco2002}
\bibinfo{author}{Cocco\xfnm[ M.]}, \bibinfo{author}{Bizzarri\xfnm[ A.]}.
\newblock \bibinfo{title}{{On the slip-weakening behavior of rate- and state
  dependent constitutive laws}}.
\newblock \bibinfo{journal}{Geophys Res Lett}
  \bibinfo{year}{2002};\bibinfo{volume}{29}(\bibinfo{number}{11}):\bibinfo{pages}{1516}.
\newblock \URLprefix \url{http://doi.wiley.com/10.1029/2001GL013999}.
  \DOIprefix\doi{10.1029/2001GL013999}.
\bibitem[{Das(2003)}]{Das2003}
\bibinfo{author}{Das\xfnm[ S.]}.
\newblock \bibinfo{title}{{Dynamic fracture mechanics in the study of the
  earthquake rupturing process: theory and observation}}.
\newblock \bibinfo{journal}{J Mech Phys Solids}
  \bibinfo{year}{2003};\bibinfo{volume}{51}(\bibinfo{number}{11-12}):\bibinfo{pages}{1939--1955}.
\newblock \URLprefix
  \url{https://linkinghub.elsevier.com/retrieve/pii/S0022509603001443}.
  \DOIprefix\doi{10.1016/j.jmps.2003.09.025}.
\bibitem[{Dieterich(2007)}]{Dietrich2007}
\bibinfo{author}{Dieterich\xfnm[ J.H.]}.
\newblock \bibinfo{title}{{Applications of rate-and state-dependent friction to
  models of fault slip and earthquake occurrence}}.
\newblock \bibinfo{journal}{Treatise Geophys}
  \bibinfo{year}{2007};\bibinfo{volume}{4}:\bibinfo{pages}{107--129}.
\newblock \DOIprefix\doi{10.1073/pnas.93.9.3787}.
\bibitem[{Freund(1998)}]{Freund1998}
\bibinfo{author}{Freund\xfnm[ L.B.]}.
\newblock \bibinfo{title}{{Dynamic Fracture Mechanics}}.
\newblock \bibinfo{address}{Cambridge}: \bibinfo{publisher}{Cambridge
  university press}, \bibinfo{year}{1998}.
\bibitem[{Geubelle and Rice(1995)}]{Geubelle1995}
\bibinfo{author}{Geubelle\xfnm[ P.]}, \bibinfo{author}{Rice\xfnm[ J.R.]}.
\newblock \bibinfo{title}{{A spectral method for three-dimensional
  elastodynamic fracture problems}}.
\newblock \bibinfo{journal}{J Mech Phys Solids}
  \bibinfo{year}{1995};\bibinfo{volume}{43}(\bibinfo{number}{11}):\bibinfo{pages}{1791--1824}.
\newblock \URLprefix
  \url{http://linkinghub.elsevier.com/retrieve/pii/002250969500043I}.
  \DOIprefix\doi{10.1016/0022-5096(95)00043-I}.
\bibitem[{Irwin(1957)}]{Irwin1957}
\bibinfo{author}{Irwin\xfnm[ G.R.]}.
\newblock \bibinfo{title}{{Analysis of stresses and strains near the end of a
  crack traversing a plate}}.
\newblock \bibinfo{journal}{J Appl Mech}
  \bibinfo{year}{1957};\bibinfo{volume}{24}:\bibinfo{pages}{361--364}.
\bibitem[{Kammer et~al.(2015)Kammer, Radiguet, Ampuero and
  Molinari}]{Kammer2015}
\bibinfo{author}{Kammer\xfnm[ D.S.]}, \bibinfo{author}{Radiguet\xfnm[ M.]},
  \bibinfo{author}{Ampuero\xfnm[ J.P.]}, \bibinfo{author}{Molinari\xfnm[
  J.F.]}.
\newblock \bibinfo{title}{{Linear Elastic Fracture Mechanics Predicts the
  Propagation Distance of Frictional Slip}}.
\newblock \bibinfo{journal}{Tribol Lett}
  \bibinfo{year}{2015};\bibinfo{volume}{57}(\bibinfo{number}{3}):\bibinfo{pages}{23}.
\newblock \URLprefix \url{http://link.springer.com/10.1007/s11249-014-0451-8}.
  \DOIprefix\doi{10.1007/s11249-014-0451-8}.
  \href{http://arxiv.org/abs/1408.4413}{\tt arXiv:1408.4413}.
\bibitem[{Kanamori and Heaton(2000)}]{kanamori2000}
\bibinfo{author}{Kanamori\xfnm[ H.]}, \bibinfo{author}{Heaton\xfnm[ T.H.]}.
\newblock \bibinfo{title}{{Microscopic and macroscopic physics of
  earthquakes}}.
\newblock In: \bibinfo{booktitle}{Geocomplexity Phys. Earthquakes}.
  \bibinfo{publisher}{American Geophysical Union (AGU)}; \bibinfo{year}{2000}.
  p. \bibinfo{pages}{147--163}.
\newblock \URLprefix
  \url{http://www.agu.org/books/gm/v120/GM120p0147/GM120p0147.shtml}.
  \DOIprefix\doi{10.1029/GM120p0147}.
\bibitem[{Lu et~al.(2010{\natexlab{a}})Lu, Lapusta and Rosakis}]{Lu2010a}
\bibinfo{author}{Lu\xfnm[ X.]}, \bibinfo{author}{Lapusta\xfnm[ N.]},
  \bibinfo{author}{Rosakis\xfnm[ A.J.]}.
\newblock \bibinfo{title}{{Pulse-like and crack-like dynamic shear ruptures on
  frictional interfaces: experimental evidence, numerical modeling, and
  implications}}.
\newblock \bibinfo{journal}{Int J Fract}
  \bibinfo{year}{2010}{\natexlab{a}};\bibinfo{volume}{163}(\bibinfo{number}{1-2}):\bibinfo{pages}{27--39}.
\newblock \URLprefix \url{http://link.springer.com/10.1007/s10704-010-9479-4}.
  \DOIprefix\doi{10.1007/s10704-010-9479-4}.
\bibitem[{Lu et~al.(2010{\natexlab{b}})Lu, Rosakis and Lapusta}]{Lu2010}
\bibinfo{author}{Lu\xfnm[ X.]}, \bibinfo{author}{Rosakis\xfnm[ A.J.]},
  \bibinfo{author}{Lapusta\xfnm[ N.]}.
\newblock \bibinfo{title}{{Rupture modes in laboratory earthquakes: Effect of
  fault prestress and nucleation conditions}}.
\newblock \bibinfo{journal}{J Geophys Res Solid Earth}
  \bibinfo{year}{2010}{\natexlab{b}};\bibinfo{volume}{115}(\bibinfo{number}{12}):\bibinfo{pages}{1--25}.
\newblock \DOIprefix\doi{10.1029/2009JB006833}.
\bibitem[{Marone(1998{\natexlab{a}})}]{Marone1998a}
\bibinfo{author}{Marone\xfnm[ C.]}.
\newblock \bibinfo{title}{{Laboratoty-derived friction laws and their
  application to seismic faulting}}.
\newblock \bibinfo{journal}{Annu Rev Earth Planet Sci}
  \bibinfo{year}{1998}{\natexlab{a}};\bibinfo{volume}{26}(\bibinfo{number}{1}):\bibinfo{pages}{643--696}.
\newblock \URLprefix
  \url{http://www.annualreviews.org/doi/abs/10.1146/annurev.earth.26.1.643}.
  \DOIprefix\doi{10.1146/annurev.earth.26.1.643}.
\bibitem[{Marone(1998{\natexlab{b}})}]{Marone1998}
\bibinfo{author}{Marone\xfnm[ C.]}.
\newblock \bibinfo{title}{{The effect of loading rate on static friction and
  the rate of fault healing during the earthquake cycle}}.
\newblock \bibinfo{journal}{Nature}
  \bibinfo{year}{1998}{\natexlab{b}};\bibinfo{volume}{391}(\bibinfo{number}{6662}):\bibinfo{pages}{69--72}.
\newblock \URLprefix
  \url{http://www.nature.com/nature/journal/v391/n6662/abs/391069a0.html}.
  \DOIprefix\doi{10.1038/34157}.
\bibitem[{Morrissey and Geubelle(1997)}]{Morrissey1997}
\bibinfo{author}{Morrissey\xfnm[ J.W.]}, \bibinfo{author}{Geubelle\xfnm[
  P.H.]}.
\newblock \bibinfo{title}{{A numerical scheme for mode III dynamic fracture
  problems}}.
\newblock \bibinfo{journal}{Int J Numer Methods Eng}
  \bibinfo{year}{1997};\bibinfo{volume}{40}(\bibinfo{number}{7}):\bibinfo{pages}{1181--1196}.
\newblock \URLprefix
  \url{https://doi.org/10.1002/(SICI)1097-0207(19970415)40:7{\%}3C1181::AID-NME108{\%}3E3.0.CO;2-X}.
  \DOIprefix\doi{10.1002/(SICI)1097-0207(19970415)40:7<1181::AID-NME108>3.0.CO;2-X}.
\bibitem[{Nagata et~al.(2012)Nagata, Nakatani and Yoshida}]{Nagata2012}
\bibinfo{author}{Nagata\xfnm[ K.]}, \bibinfo{author}{Nakatani\xfnm[ M.]},
  \bibinfo{author}{Yoshida\xfnm[ S.]}.
\newblock \bibinfo{title}{{A revised rate- and state-dependent friction law
  obtained by constraining constitutive and evolution laws separately with
  laboratory data}}.
\newblock \bibinfo{journal}{J Geophys Res Solid Earth}
  \bibinfo{year}{2012};\bibinfo{volume}{117}(\bibinfo{number}{B2}):\bibinfo{pages}{B02314}.
\newblock \URLprefix \url{http://doi.wiley.com/10.1029/2011JB008818}.
  \DOIprefix\doi{10.1029/2011JB008818}.
\bibitem[{Nakatani(2001)}]{Nakatani2001}
\bibinfo{author}{Nakatani\xfnm[ M.]}.
\newblock \bibinfo{title}{{Conceptual and physical clarification of rate and
  state friction: Frictional sliding as a thermally activated rheology}}.
\newblock \bibinfo{journal}{J Geophys Res Solid Earth}
  \bibinfo{year}{2001};\bibinfo{volume}{106}(\bibinfo{number}{B7}):\bibinfo{pages}{13347--13380}.
\newblock \URLprefix \url{http://doi.wiley.com/10.1029/2000JB900453}.
  \DOIprefix\doi{10.1029/2000JB900453}.
\bibitem[{Nielsen et~al.(2016)Nielsen, Spagnuolo, Smith, Violay, {Di Toro} and
  Bistacchi}]{Nielsen2016}
\bibinfo{author}{Nielsen\xfnm[ S.B.]}, \bibinfo{author}{Spagnuolo\xfnm[ E.]},
  \bibinfo{author}{Smith\xfnm[ S.A.F.]}, \bibinfo{author}{Violay\xfnm[ M.]},
  \bibinfo{author}{{Di Toro}\xfnm[ G.]}, \bibinfo{author}{Bistacchi\xfnm[ A.]}.
\newblock \bibinfo{title}{{Scaling in natural and laboratory earthquakes}}.
\newblock \bibinfo{journal}{Geophys Res Lett}
  \bibinfo{year}{2016};\bibinfo{volume}{43}(\bibinfo{number}{4}):\bibinfo{pages}{1504--1510}.
\newblock \URLprefix \url{http://doi.wiley.com/10.1002/2015GL067490}.
  \DOIprefix\doi{10.1002/2015GL067490}.
\bibitem[{Noda et~al.(2013)Noda, Lapusta and Kanamori}]{Noda2013a}
\bibinfo{author}{Noda\xfnm[ H.]}, \bibinfo{author}{Lapusta\xfnm[ N.]},
  \bibinfo{author}{Kanamori\xfnm[ H.]}.
\newblock \bibinfo{title}{{Comparison of average stress drop measures for
  ruptures with heterogeneous stress change and implications for earthquake
  physics}}.
\newblock \bibinfo{journal}{Geophys J Int}
  \bibinfo{year}{2013};\bibinfo{volume}{193}(\bibinfo{number}{3}):\bibinfo{pages}{1691--1712}.
\newblock \URLprefix
  \url{http://gji.oxfordjournals.org/cgi/doi/10.1093/gji/ggt074}.
  \DOIprefix\doi{10.1093/gji/ggt074}.
\bibitem[{Ohnaka(2013)}]{Ohnaka2013}
\bibinfo{author}{Ohnaka\xfnm[ M.]}.
\newblock \bibinfo{title}{{The physics of rock failure and earthquakes}}.
\newblock \bibinfo{publisher}{Cambridge University Press},
  \bibinfo{year}{2013}.
\bibitem[{Rhee et~al.(1991)Rhee, Jacko and Tsang}]{Rhee1991}
\bibinfo{author}{Rhee\xfnm[ S.]}, \bibinfo{author}{Jacko\xfnm[ M.]},
  \bibinfo{author}{Tsang\xfnm[ P.]}.
\newblock \bibinfo{title}{{The role of friction film in friction, wear and
  noise of automotive brakes}}.
\newblock \bibinfo{journal}{Wear}
  \bibinfo{year}{1991};\bibinfo{volume}{146}(\bibinfo{number}{1}):\bibinfo{pages}{89--97}.
\newblock \URLprefix
  \url{http://linkinghub.elsevier.com/retrieve/pii/004316489190226K}.
  \DOIprefix\doi{10.1016/0043-1648(91)90226-K}.
\bibitem[{Rice(2006)}]{Rice2006}
\bibinfo{author}{Rice\xfnm[ J.R.]}.
\newblock \bibinfo{title}{{Heating and weakening of faults during earthquake
  slip}}.
\newblock \bibinfo{journal}{J Geophys Res Solid Earth}
  \bibinfo{year}{2006};\bibinfo{volume}{111}(\bibinfo{number}{B5}):\bibinfo{pages}{B05311}.
\newblock \URLprefix \url{http://doi.wiley.com/10.1029/2005JB004006}.
  \DOIprefix\doi{10.1029/2005JB004006}.
\bibitem[{Rice and Ruina(1983)}]{Rice1983}
\bibinfo{author}{Rice\xfnm[ J.R.]}, \bibinfo{author}{Ruina\xfnm[ A.L.]}.
\newblock \bibinfo{title}{{Stability of Steady Frictional Slipping}}.
\newblock \bibinfo{journal}{J Appl Mech}
  \bibinfo{year}{1983};\bibinfo{volume}{50}(\bibinfo{number}{2}):\bibinfo{pages}{343--349}.
\newblock \URLprefix
  \url{http://appliedmechanics.asmedigitalcollection.asme.org/article.aspx?articleid=1406945}.
  \DOIprefix\doi{10.1115/1.3167042}.
\bibitem[{Rubin and Ampuero(2005)}]{Rubin2005}
\bibinfo{author}{Rubin\xfnm[ A.M.]}, \bibinfo{author}{Ampuero\xfnm[ J.P.]}.
\newblock \bibinfo{title}{{Earthquake nucleation on (aging) rate and state
  faults}}.
\newblock \bibinfo{journal}{J Geophys Res Solid Earth}
  \bibinfo{year}{2005};\bibinfo{volume}{110}(\bibinfo{number}{B11}):\bibinfo{pages}{B11312}.
\newblock \URLprefix \url{http://doi.wiley.com/10.1029/2005JB003686}.
  \DOIprefix\doi{10.1029/2005JB003686}.
\bibitem[{Rubino et~al.(2017)Rubino, Rosakis and Lapusta}]{Rubino2017}
\bibinfo{author}{Rubino\xfnm[ V.]}, \bibinfo{author}{Rosakis\xfnm[ A.J.]},
  \bibinfo{author}{Lapusta\xfnm[ N.]}.
\newblock \bibinfo{title}{{Understanding dynamic friction through spontaneously
  evolving laboratory earthquakes}}.
\newblock \bibinfo{journal}{Nat Commun}
  \bibinfo{year}{2017};\bibinfo{volume}{8}(\bibinfo{number}{7260}):\bibinfo{pages}{15991}.
\newblock \URLprefix \url{http://www.nature.com/doifinder/10.1038/ncomms15991}.
  \DOIprefix\doi{10.1038/ncomms15991}.
\bibitem[{Scholz(2002)}]{Scholz2002}
\bibinfo{author}{Scholz\xfnm[ C.H.]}.
\newblock \bibinfo{title}{{The mechanics of earthquakes and faulting}}.
\newblock \bibinfo{publisher}{Cambridge university press},
  \bibinfo{year}{2002}.
\bibitem[{Svetlizky et~al.(2017)Svetlizky, Bayart, Cohen and
  Fineberg}]{Svetlizky2017a}
\bibinfo{author}{Svetlizky\xfnm[ I.]}, \bibinfo{author}{Bayart\xfnm[ E.]},
  \bibinfo{author}{Cohen\xfnm[ G.]}, \bibinfo{author}{Fineberg\xfnm[ J.]}.
\newblock \bibinfo{title}{{Frictional Resistance within the Wake of Frictional
  Rupture Fronts}}.
\newblock \bibinfo{journal}{Phys Rev Lett}
  \bibinfo{year}{2017};\bibinfo{volume}{118}(\bibinfo{number}{23}):\bibinfo{pages}{234301}.
\newblock \URLprefix
  \url{http://link.aps.org/doi/10.1103/PhysRevLett.118.234301}.
  \DOIprefix\doi{10.1103/PhysRevLett.118.234301}.
\bibitem[{Svetlizky et~al.(2019)Svetlizky, Bayart and Fineberg}]{Svetlizky2019}
\bibinfo{author}{Svetlizky\xfnm[ I.]}, \bibinfo{author}{Bayart\xfnm[ E.]},
  \bibinfo{author}{Fineberg\xfnm[ J.]}.
\newblock \bibinfo{title}{{Brittle Fracture Theory Describes the Onset of
  Frictional Motion}}.
\newblock \bibinfo{journal}{Annu Rev Condens Matter Phys}
  \bibinfo{year}{2019};\bibinfo{volume}{10}(\bibinfo{number}{1}):\bibinfo{pages}{031218--013327}.
\newblock \URLprefix
  \url{https://www.annualreviews.org/doi/10.1146/annurev-conmatphys-031218-013327}.
  \DOIprefix\doi{10.1146/annurev-conmatphys-031218-013327}.
\bibitem[{Svetlizky and Fineberg(2014)}]{Svetlizky2014}
\bibinfo{author}{Svetlizky\xfnm[ I.]}, \bibinfo{author}{Fineberg\xfnm[ J.]}.
\newblock \bibinfo{title}{{Classical shear cracks drive the onset of dry
  frictional motion}}.
\newblock \bibinfo{journal}{Nature}
  \bibinfo{year}{2014};\bibinfo{volume}{509}(\bibinfo{number}{7499}):\bibinfo{pages}{205--208}.
\newblock \URLprefix \url{https://www.nature.com/articles/nature13202}.
  \DOIprefix\doi{10.1038/nature13202}.
\bibitem[{Svetlizky et~al.(2016)Svetlizky, {Pino Mu{\~{n}}oz}, Radiguet,
  Kammer, Molinari and Fineberg}]{Svetlizky2016}
\bibinfo{author}{Svetlizky\xfnm[ I.]}, \bibinfo{author}{{Pino
  Mu{\~{n}}oz}\xfnm[ D.]}, \bibinfo{author}{Radiguet\xfnm[ M.]},
  \bibinfo{author}{Kammer\xfnm[ D.S.]}, \bibinfo{author}{Molinari\xfnm[ J.F.]},
  \bibinfo{author}{Fineberg\xfnm[ J.]}.
\newblock \bibinfo{title}{{Properties of the shear stress peak radiated ahead
  of rapidly accelerating rupture fronts that mediate frictional slip}}.
\newblock \bibinfo{journal}{Proc Natl Acad Sci}
  \bibinfo{year}{2016};\bibinfo{volume}{113}(\bibinfo{number}{3}):\bibinfo{pages}{542--547}.
\newblock \URLprefix
  \url{http://www.pnas.org/lookup/doi/10.1073/pnas.1517545113}.
  \DOIprefix\doi{10.1073/pnas.1517545113}.
\bibitem[{Tinti et~al.(2005)Tinti, Spudich and Cocco}]{Tinti2005}
\bibinfo{author}{Tinti\xfnm[ E.]}, \bibinfo{author}{Spudich\xfnm[ P.]},
  \bibinfo{author}{Cocco\xfnm[ M.]}.
\newblock \bibinfo{title}{{Earthquake fracture energy inferred from kinematic
  rupture models on extended faults}}.
\newblock \bibinfo{journal}{J Geophys Res}
  \bibinfo{year}{2005};\bibinfo{volume}{110}(\bibinfo{number}{B12}):\bibinfo{pages}{B12303}.
\newblock \URLprefix \url{http://doi.wiley.com/10.1029/2005JB003644}.
  \DOIprefix\doi{10.1029/2005JB003644}.
\bibitem[{Viesca and Garagash(2015)}]{Viesca2015}
\bibinfo{author}{Viesca\xfnm[ R.C.]}, \bibinfo{author}{Garagash\xfnm[ D.I.]}.
\newblock \bibinfo{title}{{Ubiquitous weakening of faults due to thermal
  pressurization}}.
\newblock \bibinfo{journal}{Nat Geosci}
  \bibinfo{year}{2015};\bibinfo{volume}{8}(\bibinfo{number}{11}):\bibinfo{pages}{875--879}.
\newblock \URLprefix \url{https://www.nature.com/articles/ngeo2554}.
  \DOIprefix\doi{10.1038/ngeo2554}.
\end{thebibliography}

\begin{thebibliography}{13}%
\makeatletter
\providecommand \@ifxundefined [1]{%
 \@ifx{#1\undefined}
}%
\providecommand \@ifnum [1]{%
 \ifnum #1\expandafter \@firstoftwo
 \else \expandafter \@secondoftwo
 \fi
}%
\providecommand \@ifx [1]{%
 \ifx #1\expandafter \@firstoftwo
 \else \expandafter \@secondoftwo
 \fi
}%
\providecommand \natexlab [1]{#1}%
\providecommand \enquote  [1]{``#1''}%
\providecommand \bibnamefont  [1]{#1}%
\providecommand \bibfnamefont [1]{#1}%
\providecommand \citenamefont [1]{#1}%
\providecommand \href@noop [0]{\@secondoftwo}%
\providecommand \href [0]{\begingroup \@sanitize@url \@href}%
\providecommand \@href[1]{\@@startlink{#1}\@@href}%
\providecommand \@@href[1]{\endgroup#1\@@endlink}%
\providecommand \@sanitize@url [0]{\catcode `\\12\catcode `\$12\catcode
  `\&12\catcode `\#12\catcode `\^12\catcode `\_12\catcode `\%12\relax}%
\providecommand \@@startlink[1]{}%
\providecommand \@@endlink[0]{}%
\providecommand \url  [0]{\begingroup\@sanitize@url \@url }%
\providecommand \@url [1]{\endgroup\@href {#1}{\urlprefix }}%
\providecommand \urlprefix  [0]{URL }%
\providecommand \Eprint [0]{\href }%
\providecommand \doibase [0]{https://doi.org/}%
\providecommand \selectlanguage [0]{\@gobble}%
\providecommand \bibinfo  [0]{\@secondoftwo}%
\providecommand \bibfield  [0]{\@secondoftwo}%
\providecommand \translation [1]{[#1]}%
\providecommand \BibitemOpen [0]{}%
\providecommand \bibitemStop [0]{}%
\providecommand \bibitemNoStop [0]{.\EOS\space}%
\providecommand \EOS [0]{\spacefactor3000\relax}%
\providecommand \BibitemShut  [1]{\csname bibitem#1\endcsname}%
\let\auto@bib@innerbib\@empty
\bibitem [{\citenamefont {Barras}\ \emph {et~al.}(2019)\citenamefont {Barras},
  \citenamefont {Aldam}, \citenamefont {Roch}, \citenamefont {Brener},
  \citenamefont {Bouchbinder},\ and\ \citenamefont {Molinari}}]{PartI}%
  \BibitemOpen
  \bibfield  {author} {\bibinfo {author} {\bibfnamefont {F.}~\bibnamefont
  {Barras}}, \bibinfo {author} {\bibfnamefont {M.}~\bibnamefont {Aldam}},
  \bibinfo {author} {\bibfnamefont {T.}~\bibnamefont {Roch}}, \bibinfo {author}
  {\bibfnamefont {E.~A.}\ \bibnamefont {Brener}}, \bibinfo {author}
  {\bibfnamefont {E.}~\bibnamefont {Bouchbinder}},\ and\ \bibinfo {author}
  {\bibfnamefont {J.-F.}\ \bibnamefont {Molinari}},\ }\bibfield  {title}
  {\bibinfo {title} {{The emergence of crack-like behavior of frictional
  rupture: The origin of stress drops}},\ }\href@noop {} {\bibfield  {journal}
  {\bibinfo  {journal} {To appear in Physical Review X}\ } (\bibinfo {year} {2019})}\BibitemShut
  {NoStop}%
\bibitem [{\citenamefont {Brener}\ \emph {et~al.}(2018)\citenamefont {Brener},
  \citenamefont {Aldam}, \citenamefont {Barras}, \citenamefont {Molinari},\
  and\ \citenamefont {Bouchbinder}}]{Brener2018}%
  \BibitemOpen
  \bibfield  {author} {\bibinfo {author} {\bibfnamefont {E.~A.}\ \bibnamefont
  {Brener}}, \bibinfo {author} {\bibfnamefont {M.}~\bibnamefont {Aldam}},
  \bibinfo {author} {\bibfnamefont {F.}~\bibnamefont {Barras}}, \bibinfo
  {author} {\bibfnamefont {J.-F.}\ \bibnamefont {Molinari}},\ and\ \bibinfo
  {author} {\bibfnamefont {E.}~\bibnamefont {Bouchbinder}},\ }\bibfield
  {title} {\bibinfo {title} {{Unstable Slip Pulses and Earthquake Nucleation as
  a Nonequilibrium First-Order Phase Transition}},\ }\href
  {https://doi.org/10.1103/PhysRevLett.121.234302} {\bibfield  {journal}
  {\bibinfo  {journal} {Phys. Rev. Lett.}\ }\textbf {\bibinfo {volume} {121}},\
  \bibinfo {pages} {234302} (\bibinfo {year} {2018})}\BibitemShut {NoStop}%
\bibitem [{\citenamefont {Dugdale}(1960)}]{Dugdale1960}%
  \BibitemOpen
  \bibfield  {author} {\bibinfo {author} {\bibfnamefont {D.}~\bibnamefont
  {Dugdale}},\ }\bibfield  {title} {\bibinfo {title} {{Yielding of steel sheets
  containing slits}},\ }\href {https://doi.org/10.1016/0022-5096(60)90013-2}
  {\bibfield  {journal} {\bibinfo  {journal} {J. Mech. Phys. Solids}\ }\textbf
  {\bibinfo {volume} {8}},\ \bibinfo {pages} {100} (\bibinfo {year}
  {1960})}\BibitemShut {NoStop}%
\bibitem [{\citenamefont {Barenblatt}(1962)}]{Barenblatt1962}%
  \BibitemOpen
  \bibfield  {author} {\bibinfo {author} {\bibfnamefont {G.}~\bibnamefont
  {Barenblatt}},\ }\bibfield  {title} {\bibinfo {title} {{The Mathematical
  Theory of Equilibrium Cracks in Brittle Fracture}},\ }\href
  {https://doi.org/10.1016/S0065-2156(08)70121-2} {\bibfield  {journal}
  {\bibinfo  {journal} {Adv. Appl. Mech.}\ }\textbf {\bibinfo {volume} {7}},\
  \bibinfo {pages} {55} (\bibinfo {year} {1962})}\BibitemShut {NoStop}%
\bibitem [{\citenamefont {Breitenfeld}\ and\ \citenamefont
  {Geubelle}(1998)}]{Breitenfeld1998}%
  \BibitemOpen
  \bibfield  {author} {\bibinfo {author} {\bibfnamefont {M.~S.}\ \bibnamefont
  {Breitenfeld}}\ and\ \bibinfo {author} {\bibfnamefont {P.~H.}\ \bibnamefont
  {Geubelle}},\ }\bibfield  {title} {\bibinfo {title} {{Numerical analysis of
  dynamic debonding under 2D in-plane and 3D loading}},\ }\href
  {https://doi.org/10.1023/A:1007535703095} {\bibfield  {journal} {\bibinfo
  {journal} {Int. J. Fract.}\ }\textbf {\bibinfo {volume} {93}},\ \bibinfo
  {pages} {13} (\bibinfo {year} {1998})}\BibitemShut {NoStop}%
\bibitem [{\citenamefont {Barras}\ \emph {et~al.}(2014)\citenamefont {Barras},
  \citenamefont {Kammer}, \citenamefont {Geubelle},\ and\ \citenamefont
  {Molinari}}]{Barras2014}%
  \BibitemOpen
  \bibfield  {author} {\bibinfo {author} {\bibfnamefont {F.}~\bibnamefont
  {Barras}}, \bibinfo {author} {\bibfnamefont {D.~S.}\ \bibnamefont {Kammer}},
  \bibinfo {author} {\bibfnamefont {P.~H.}\ \bibnamefont {Geubelle}},\ and\
  \bibinfo {author} {\bibfnamefont {J.-F.}\ \bibnamefont {Molinari}},\
  }\bibfield  {title} {\bibinfo {title} {{A study of frictional contact in
  dynamic fracture along bimaterial interfaces}},\ }\href
  {https://doi.org/10.1007/s10704-014-9967-z} {\bibfield  {journal} {\bibinfo
  {journal} {Int. J. Fract.}\ }\textbf {\bibinfo {volume} {189}},\ \bibinfo
  {pages} {149} (\bibinfo {year} {2014})}\BibitemShut {NoStop}%
\bibitem [{\citenamefont {Andrews}(1976)}]{Andrews1976}%
  \BibitemOpen
  \bibfield  {author} {\bibinfo {author} {\bibfnamefont {D.~J.}\ \bibnamefont
  {Andrews}},\ }\bibfield  {title} {\bibinfo {title} {{Rupture propagation with
  finite stress in antiplane strain}},\ }\href
  {https://doi.org/10.1029/JB081i020p03575} {\bibfield  {journal} {\bibinfo
  {journal} {J. Geophys. Res.}\ }\textbf {\bibinfo {volume} {81}},\ \bibinfo
  {pages} {3575} (\bibinfo {year} {1976})}\BibitemShut {NoStop}%
\bibitem [{\citenamefont {Freund}(1998)}]{Freund1998}%
  \BibitemOpen
  \bibfield  {author} {\bibinfo {author} {\bibfnamefont {L.~B.}\ \bibnamefont
  {Freund}},\ }\href@noop {} {\emph {\bibinfo {title} {{Dynamic Fracture
  Mechanics}}}}\ (\bibinfo  {publisher} {Cambridge university press},\ \bibinfo
  {address} {Cambridge},\ \bibinfo {year} {1998})\BibitemShut {NoStop}%
\bibitem [{\citenamefont {Jones}\ \emph {et~al.}(2001)\citenamefont {Jones},
  \citenamefont {Oliphant},\ and\ \citenamefont {Peterson}}]{Jones2001}%
  \BibitemOpen
  \bibfield  {author} {\bibinfo {author} {\bibfnamefont {E.}~\bibnamefont
  {Jones}}, \bibinfo {author} {\bibfnamefont {T.}~\bibnamefont {Oliphant}},\
  and\ \bibinfo {author} {\bibfnamefont {P.}~\bibnamefont {Peterson}},\
  }\bibfield  {title} {\bibinfo {title} {{SciPy: Open source scientific tools
  for Python}}}\href@noop {} {\ (\bibinfo {year} {2001})}\BibitemShut
  {NoStop}%
\bibitem [{\citenamefont {Palmer}\ and\ \citenamefont
  {Rice}(1973)}]{Palmer1973}%
  \BibitemOpen
  \bibfield  {author} {\bibinfo {author} {\bibfnamefont {A.~C.}\ \bibnamefont
  {Palmer}}\ and\ \bibinfo {author} {\bibfnamefont {J.~R.}\ \bibnamefont
  {Rice}},\ }\bibfield  {title} {\bibinfo {title} {{The Growth of Slip Surfaces
  in the Progressive Failure of Over-Consolidated Clay}},\ }\href
  {https://doi.org/10.1098/rspa.1973.0040} {\bibfield  {journal} {\bibinfo
  {journal} {Proc. R. Soc. A Math. Phys. Eng. Sci.}\ }\textbf {\bibinfo
  {volume} {332}},\ \bibinfo {pages} {527} (\bibinfo {year}
  {1973})}\BibitemShut {NoStop}%
\bibitem [{\citenamefont {Baumberger}\ and\ \citenamefont
  {Caroli}(2006)}]{Baumberger2006}%
  \BibitemOpen
  \bibfield  {author} {\bibinfo {author} {\bibfnamefont {T.}~\bibnamefont
  {Baumberger}}\ and\ \bibinfo {author} {\bibfnamefont {C.}~\bibnamefont
  {Caroli}},\ }\bibfield  {title} {\bibinfo {title} {{Solid friction from
  stick-slip down to pinning and aging}},\ }\href
  {https://doi.org/10.1080/00018730600732186} {\bibfield  {journal} {\bibinfo
  {journal} {Adv. Phys.}\ }\textbf {\bibinfo {volume} {55}},\ \bibinfo {pages}
  {279} (\bibinfo {year} {2006})}\BibitemShut {NoStop}%
\bibitem [{\citenamefont {Baumberger}\ and\ \citenamefont
  {Berthoud}(1999)}]{Baumberger1999}%
  \BibitemOpen
  \bibfield  {author} {\bibinfo {author} {\bibfnamefont {T.}~\bibnamefont
  {Baumberger}}\ and\ \bibinfo {author} {\bibfnamefont {P.}~\bibnamefont
  {Berthoud}},\ }\bibfield  {title} {\bibinfo {title} {{Physical analysis of
  the state- and rate-dependent friction law. II. Dynamic friction}},\ }\href
  {https://doi.org/10.1103/PhysRevB.60.3928} {\bibfield  {journal} {\bibinfo
  {journal} {Phys. Rev. B}\ }\textbf {\bibinfo {volume} {60}},\ \bibinfo
  {pages} {3928} (\bibinfo {year} {1999})}\BibitemShut {NoStop}%
\bibitem [{\citenamefont {Bar-Sinai}\ \emph {et~al.}(2014)\citenamefont
  {Bar-Sinai}, \citenamefont {Spatschek}, \citenamefont {Brener},\ and\
  \citenamefont {Bouchbinder}}]{Bar-Sinai2014}%
  \BibitemOpen
  \bibfield  {author} {\bibinfo {author} {\bibfnamefont {Y.}~\bibnamefont
  {Bar-Sinai}}, \bibinfo {author} {\bibfnamefont {R.}~\bibnamefont
  {Spatschek}}, \bibinfo {author} {\bibfnamefont {E.~A.}\ \bibnamefont
  {Brener}},\ and\ \bibinfo {author} {\bibfnamefont {E.}~\bibnamefont
  {Bouchbinder}},\ }\bibfield  {title} {\bibinfo {title} {{On the
  velocity-strengthening behavior of dry friction}},\ }\href
  {https://doi.org/10.1002/2013JB010586} {\bibfield  {journal} {\bibinfo
  {journal} {J. Geophys. Res. Solid Earth}\ }\textbf {\bibinfo {volume}
  {119}},\ \bibinfo {pages} {1738} (\bibinfo {year} {2014})}\BibitemShut
  {NoStop}%
\bibitem [{\citenamefont {Molinari}\ and\ \citenamefont
  {Perfettini}(2019)}]{Molinari2019}%
  \BibitemOpen
  \bibfield  {author} {\bibinfo {author} {\bibfnamefont {A.}~\bibnamefont
  {Molinari}}\ and\ \bibinfo {author} {\bibfnamefont {H.}~\bibnamefont
  {Perfettini}},\ }\bibfield  {title} {\bibinfo {title} {{Fundamental aspects
  of a new micromechanical model of rate and state friction}},\ }\href
  {https://doi.org/10.1016/j.jmps.2018.10.002} {\bibfield  {journal} {\bibinfo
  {journal} {J. Mech. Phys. Solids}\ }\textbf {\bibinfo {volume} {124}},\
  \bibinfo {pages} {63} (\bibinfo {year} {2019})}\BibitemShut {NoStop}%
\bibitem [{\citenamefont {Cocco}\ and\ \citenamefont
  {Bizzarri}(2002)}]{Cocco2002}%
  \BibitemOpen
  \bibfield  {author} {\bibinfo {author} {\bibfnamefont {M.}~\bibnamefont
  {Cocco}}\ and\ \bibinfo {author} {\bibfnamefont {A.}~\bibnamefont
  {Bizzarri}},\ }\bibfield  {title} {\bibinfo {title} {{On the slip-weakening
  behavior of rate- and state dependent constitutive laws}},\ }\href
  {https://doi.org/10.1029/2001GL013999} {\bibfield  {journal} {\bibinfo
  {journal} {Geophys. Res. Lett.}\ }\textbf {\bibinfo {volume} {29}},\ \bibinfo
  {pages} {1516} (\bibinfo {year} {2002})}\BibitemShut {NoStop}%
\bibitem [{\citenamefont {Bizzarri}\ and\ \citenamefont
  {Cocco}(2003)}]{Bizzarri2003}%
  \BibitemOpen
  \bibfield  {author} {\bibinfo {author} {\bibfnamefont {A.}~\bibnamefont
  {Bizzarri}}\ and\ \bibinfo {author} {\bibfnamefont {M.}~\bibnamefont
  {Cocco}},\ }\bibfield  {title} {\bibinfo {title} {{Slip-weakening behavior
  during the propagation of dynamic ruptures obeying rate- and state-dependent
  friction laws}},\ }\href {https://doi.org/10.1029/2002JB002198} {\bibfield
  {journal} {\bibinfo  {journal} {J. Geophys. Res.}\ }\textbf {\bibinfo
  {volume} {108}},\ \bibinfo {pages} {2373} (\bibinfo {year}
  {2003})}\BibitemShut {NoStop}%
\bibitem [{\citenamefont {Rubin}\ and\ \citenamefont
  {Ampuero}(2005)}]{Rubin2005}%
  \BibitemOpen
  \bibfield  {author} {\bibinfo {author} {\bibfnamefont {A.~M.}\ \bibnamefont
  {Rubin}}\ and\ \bibinfo {author} {\bibfnamefont {J.-P.}\ \bibnamefont
  {Ampuero}},\ }\bibfield  {title} {\bibinfo {title} {{Earthquake nucleation on
  (aging) rate and state faults}},\ }\href
  {https://doi.org/10.1029/2005JB003686} {\bibfield  {journal} {\bibinfo
  {journal} {J. Geophys. Res. Solid Earth}\ }\textbf {\bibinfo {volume}
  {110}},\ \bibinfo {pages} {B11312} (\bibinfo {year} {2005})}\BibitemShut
  {NoStop}%
\end{thebibliography}

%

\end{document}